\documentclass{aa}

\usepackage{graphicx}
\usepackage{nccmath}
\usepackage{xcolor}
\usepackage{txfonts}
\usepackage{subfigure}
\usepackage{float}
\usepackage{natbib,twoopt}
\usepackage{xspace}
\usepackage{balance}
\usepackage{upgreek}
\usepackage{multirow}
\usepackage[normalem]{ulem}
\usepackage{soul}
\usepackage[colorlinks=true, allcolors=blue]{hyperref}
\usepackage{outlines}
\usepackage{relsize}

\usepackage[flushleft]{threeparttable}

\definecolor{redUniBo}{RGB}{187, 46, 41}

\def\apjl{ApJ Lett.}
\def\apjs{ApJ Suppl.}

\def\aap{A\&A}

\bibpunct{(}{)}{;}{a}{}{,}
\DeclareGraphicsExtensions{.eps,.ps,.pdf}

\begin{document}

\defcitealias{Lesci25}{L25}
\defcitealias{Maturi25}{M25}
\defcitealias{Diemer14}{DK14}
\defcitealias{Planck18}{Planck18}

\title{AMICO galaxy clusters in KiDS-1000: Splashback radius from weak lensing and the cluster--galaxy correlation function}

\author
{G. F. Lesci\inst{\ref{1},\ref{2}}
\and C. Giocoli\inst{\ref{1},\ref{2},\ref{3}}
\and F. Marulli\inst{\ref{1},\ref{2},\ref{3}}
\and M. Romanello\inst{\ref{1},\ref{2}}
\and L. Moscardini\inst{\ref{1},\ref{2},\ref{3}}
\and M. Sereno\inst{\ref{2},\ref{3}}
\and M. Maturi\inst{\ref{5},\ref{8}}
\and \\ M. Radovich\inst{\ref{7}}
\and G. Castignani\inst{\ref{2}}
\and H. Hildebrandt\inst{\ref{bochum}}
\and L. Ingoglia\inst{\ref{1},\ref{10}}
\and E. Puddu\inst{\ref{9}}
}

\offprints{G. F. Lesci \\ \email{giorgio.lesci2@unibo.it}}

\institute{
  Dipartimento di Fisica e Astronomia ``Augusto Righi'' - Alma Mater Studiorum
  Universit\`{a} di Bologna, via Piero Gobetti 93/2, I-40129 Bologna,
  Italy\label{1}
  \and INAF - Osservatorio di Astrofisica e Scienza dello Spazio di
  Bologna, via Piero Gobetti 93/3, I-40129 Bologna, Italy\label{2}
  \and INFN - Sezione di Bologna, viale Berti Pichat 6/2, I-40127
  Bologna, Italy\label{3}
  \and Zentrum f\"ur Astronomie, Universit\"at Heidelberg, Philosophenweg 12, D-69120 Heidelberg, Germany\label{5}
  \and ITP, Universit\"at Heidelberg, Philosophenweg 16, 69120 Heidelberg, Germany\label{8}  
  \and INAF - Osservatorio Astronomico di Padova, vicolo dell'Osservatorio 5, I-35122 Padova, Italy\label{7}
  \and INAF - Istituto di Radioastronomia, Via Piero Gobetti 101, 40129 Bologna, Italy\label{10}
  \and Ruhr University Bochum, Faculty of Physics and Astronomy, Astronomical Institute (AIRUB), German Centre for Cosmological Lensing, 44780 Bochum, Germany\label{bochum}
  \and INAF - Osservatorio Astronomico di Capodimonte, Salita Moiariello 16, Napoli 80131, Italy\label{9}
}

\date{Received --; accepted --}

\abstract
   {}
   {
   We present the splashback radius analysis of the Adaptive Matched Identifier of Clustered Objects (AMICO) galaxy cluster sample in the fourth data release of the Kilo Degree Survey (KiDS). The sample contains 9049 rich galaxy clusters within $z\in[0.1,0.8]$, with shear measurements available for 8730 of them.
   }
   {
    We measured and modelled the stacked reduced shear, $g_{\rm t}$, and the cluster--galaxy correlation function, $w_{\rm cg}$, in bins of observed intrinsic richness, $\lambda^*$, and redshift, $z$. Building on the methods employed in recent cosmological analyses, we modelled the average splashback radius, $r_{\rm sp}$, of the underlying dark matter halo distribution, accounting for the known systematic uncertainties that affect measurements and theoretical models.
    }
   {
   By modelling $g_{\rm t}$ and $w_{\rm cg}$ separately in the cluster-centric radial range $R\in[0.4,5]$ $h^{-1}$Mpc, we constrain $r_{\rm sp}$, the mass accretion rate, $\Gamma$, and the relation between $\mathcal{R}_{\rm sp}\equiv r_{\rm sp}/r_{200\rm m}$ and the peak height, $\nu_{200\rm m}$, over the mass range $M_{200\rm m}\in[0.4,20]$ $10^{14}h^{-1}$M$_\odot$. The two probes provide consistent results that also agree with $\Lambda$-cold dark matter model predictions. Our $\mathcal{R}_{\rm sp}$ constraints are consistent with those from previous observations. For $g_{\rm t}$ and $w_{\rm cg}$, we achieve a precision of 14\% and 10\% per cluster stack, respectively. The higher precision of $w_{\rm cg}$, enabled by its combination with weak-lensing constraints on the mass-richness relation, highlights the complementarity of lensing and clustering in measuring $r_{\rm sp}$ and constraining the properties of the infalling material region.
   }
   {}

\keywords{galaxies: clusters: general -- galaxies: kinematics and dynamics -- cosmology: observations}

\authorrunning{G. F. Lesci et al.}

\titlerunning{AMICO clusters in KiDS-1000: Splashback radius}

\maketitle

\section{Introduction}
The mass profiles of dark matter haloes have been extensively investigated using $N$-body simulations \citep*[see e.g.][]{Einasto1965,NFW,BMO}, establishing a robust theoretical foundation for interpreting galaxy cluster observations. Precise measurements of the inner and virialised regions of clusters not only support cosmological analyses based on cluster statistics \citep{Planck_counts,Costanzi19,Marulli21,Romanello24,Seppi24,Ghirardini24,Lesci25} but also enable tests of gravity theories \citep{Pizzuti17,Cardone21,Mitchell21,Rosselli23} and constraints on dark matter self-interactions \citep{Peter13,Robertson19,Eckert22,Bhattacharyya22,Despali25}. In addition, measurements in the outer halo regime --- which is commonly referred to as the two-halo region and is dominant beyond 2--5 Mpc from cluster centres \citep{Tinker05} --- offer a complementary route to constrain cosmological parameters and cluster masses \citep{Covone14,Sereno18,Giocoli21,Ingoglia22,Eltvedt24}. \\
\indent In the transition between the virialised region and the two-halo regime, a sharp steepening in the density profile is expected. This feature, located in the proximity of the splashback radius, $r_{\rm sp}$, was initially predicted from analytical models of  spherical secondary collapse by \citet{Fillmore1984} and \citet{Bertschinger1985}, and then robustly identified in dark matter simulations \citep{Diemer14,Adhikari14,More15}. The splashback radius marks the apocentre of the first orbit of recently accreted material. As a consequence, it strongly depends on the mass accretion rate, $\Gamma$. At fixed mass, haloes with higher $\Gamma$ exhibit smaller $r_{\rm sp}$ due to the deepening of the gravitational potential well during the orbit of infalling material. More massive sub-haloes also tend to have a smaller orbit apocentre than the less massive ones. This effect is attributed to dynamical friction \citep{Diemer17,Chang18}, which slows satellites by dissipating kinetic energy, thereby bringing them deeper into the potential well. Unlike halo concentration, the location of $r_{\rm sp}$ is more sensitive to recent accretion (within the last crossing time) than to the entire assembly history \citep{Shin23}. However, the depth and width of the associated density steepening are strongly influenced by the full halo assembly history, providing complementary information on the location of the splashback radius \citep{Yu25}. \\
\indent As $r_{\rm sp}$ delineates a boundary between first-infalling matter and matter that has completed at least one passage through the central region of the halo, the mass enclosed within $r_{\rm sp}$ comprises all material accreted by a given redshift, $z$. Therefore, this mass definition remains unaffected by pseudo-evolution \citep{More15} and can potentially lead to the derivation of universal halo mass function models \citep{Diemer20,Ryu21,Ryu22}. In fact, recent studies highlight the distinction between orbiting and infalling particles as the key factor in the definition of a halo (see \citealt{Vladimir25} and references therein). Furthermore, $r_{\rm sp}$ has been shown to correlate with fundamental cosmological parameters, such as the matter density parameter at $z=0$, $\Omega_{\rm m}$ and the square root of the mass variance on a scale of 8 $h^{-1}$Mpc at $z=0$, $\sigma_8$ \citep{Diemer17,Haggar24,Mpetha24}.\\
\indent It is critical to note that the physical origin of $r_{\rm sp}$ lies in the dynamics of halo particles. \citet{Diemer17_2,Diemer20_2} show that the radius of the steepest slope, which is directly observable in the mass density profiles, is a scattered proxy of $r_{\rm sp}$ \citep[see also][]{Sun25}. This occurs because the apocentres are distributed across a wide radial range due to non-spherical halo shapes, complex accretion histories, and gravitational interactions with neighbouring haloes. Alternative halo boundary definitions have been proposed in the literature, such as the depletion radius \citep{Fong21,Zhou23}, which is about 2 times larger than $r_{\rm sp}$ and corresponds to the scale above which the matter density decreases with time. \citet{Pizzardo24} demonstrate that the inflection point in the radial velocity profile of cluster galaxies agrees with $r_{\rm sp}$ within $1\sigma$, establishing an additional connection between the splashback radius and the inner boundary of the cluster infall region. \citet{Garcia23} and \citet{Salazar25} show that a clear truncation radius is apparent in the halo–matter correlation function when particles are classified as bound or infalling based on their phase-space trajectories and accretion history.\\
\indent In this work, following the standard nomenclature in the literature, we define the radius of the steepest slope as the splashback radius. We constrained the splashback radii of the galaxy clusters detected with the Adaptive Matched Identifier of Clustered Objects \citep[AMICO;][]{Bellagamba18,Maturi19} in the fourth data release of the Kilo Degree Survey \citep[KiDS-1000;][]{Kuijken19}. The cluster sample used in this work comprises 8730 rich clusters, spanning a total mass range of $M_{200\rm m}\in[0.4,20]$ $10^{14}h^{-1}$M$_\odot$ and extending up to a redshift $z=0.8$. We measured and modelled the stacked weak-lensing profiles and cluster--galaxy correlation function of ensembles of clusters, constraining the relation between the peak height, $\nu_{200\rm m}$, and the normalised splashback radius, $\mathcal{R}_{\rm sp}\equiv r_{\rm sp}/r_{200\rm m}$, where $r_{200\rm m}$ is the radius enclosing a mass $M_{200\rm m}$, such that the corresponding mean density is 200 times the mean density of the Universe at the halo redshift. We also derived the mass accretion rate as a function of mass and redshift. Our results are competitive, in terms of statistical uncertainties, with those from previous analyses of observed cluster samples \citep{More16,Baxter17,Umetsu17,Chang18,Contigiani19,Zurcher19,Shin19,Murata20,Bianconi21,Shin21,Rana23,Joshi25}. Compared to the analogous analysis by \citet{Giocoli24}, based on AMICO detections in the third data release of KiDS \citep[KiDS-DR3;][]{deJong17}, we employed more conservative galaxy and cluster selections, and used a more extensive modelling of systematic uncertainties. Nevertheless, the precision on the splashback radius we achieve is comparable to that of \citet{Giocoli24} since the survey effective area is more than doubled compared to KiDS-DR3, and the cluster sample extends up to higher redshifts. \\
\indent The paper is organised as follows. In Sect.\ \ref{sec:dataset} we present the galaxy cluster and shear samples, and in Sect.\ \ref{sec:measure} the measurements of stacked weak-lensing profiles and the cluster--galaxy correlation function. In Sect.\ \ref{sec:modelling} we outline the theoretical framework, likelihood function, and prior distributions used in this work. We present our results in Sect.~\ref{sec:results} and our conclusions in Sect.~\ref{sec:conclusions}. \\
\indent In this work we adopted a concordance flat $\Lambda$ cold dark matter ($\Lambda$CDM) cosmological model. The base 10 logarithm is referred to as $\log$, while $\ln$ represents the natural logarithm. The statistical analyses presented in this paper were performed using the \texttt{CosmoBolognaLib}\footnote{\url{https://gitlab.com/federicomarulli/CosmoBolognaLib/}} \citep{cbl}, a set of free software C++/Python numerical libraries for cosmological calculations. The linear matter power spectrum was computed with \texttt{CAMB}\footnote{\url{https://camb.info/}} \citep{CAMB}.

\section{Dataset}\label{sec:dataset}
\begin{figure*}[t!]
\centering\includegraphics[width = \hsize, height = 4.5cm] {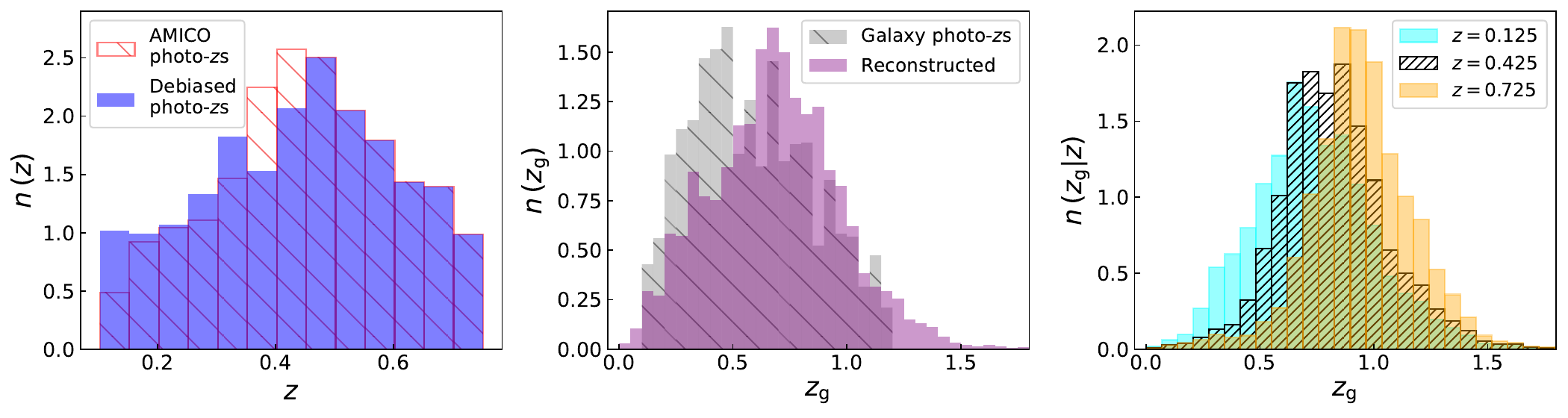}
\caption{\textit{Left}: Cluster photo-$z$ distributions as measured by AMICO (hatched red) and unbiased using a reference spectroscopic sample (blue). \textit{Middle}: Observed photo-$z$ (hatched grey) and SOM-reconstructed (purple) redshift distributions of the full galaxy sample. \textit{Right}: Examples of SOM-reconstructed background galaxy redshift distributions, given a cluster redshift of $z=0.125$ (cyan), $z=0.425$ (hatched black), $z=0.725$ (orange), derived by using the background selections by \citetalias{Lesci25}.}
\label{fig:redshift_distributions}
\end{figure*}
In our analysis we used KiDS-1000 data \citep[][]{Kuijken19}, which are based on imaging obtained with the OmegaCAM wide-field camera \citep{OmegaCAM}, operating on the VLT Survey Telescope \citep[VST;][]{VST}. The KiDS-1000 release comprises 1006 square-degree tiles, each spanning roughly 1 deg$^2$. It includes 2 arcsecond aperture photometry in the $u$, $g$, $r$, and $i$ optical bands, with 5$\sigma$ depth limits of 24.23, 25.12, 25.02, and 23.68 magnitudes, respectively. Complementary near-infrared photometry in the $Z$, $Y$, $J$, $H$, and $K_{\rm s}$ bands is also provided by the VISTA Kilo-degree INfrared Galaxy (VIKING) survey \citep[][]{VIKINGS,VISTA}, which fully overlaps the KiDS footprint. \\
\indent The galaxy cluster sample analysed in this work originates from the AMICO KiDS-1000 catalogue \citep[][M25 hereafter]{Maturi25}, constructed by applying the AMICO algorithm \citep{Bellagamba18,Maturi19} to KiDS-1000 photometric data. AMICO uses an optimal matched filter and is one of the official cluster-finding tools of the \textit{Euclid} mission \citep{Adam19}. The catalogue excludes regions affected by image artefacts and photometric issues, resulting in a clean survey area of 839 deg$^2$, with cluster detections up to $z=0.9$. Following \citet[][L25 hereafter]{Lesci25}, we restricted our analysis to clusters with ${\rm S/N}>3.5$ and redshifts $z\in[0.1,0.8]$, yielding 22\,396 detections. Additional cuts in the cluster mass proxy, namely the intrinsic richness $\lambda^*$ \citepalias[see][]{Maturi25}, yield a final cluster sample of 8730 objects, as discussed in Sect.\ \ref{sec:measure}. These selections optimise the trade-off between fit quality and sample size \citepalias{Lesci25}. Cluster redshifts have been corrected for the bias identified by \citetalias{Maturi25}, based on a comparison with spectroscopic redshifts available within the KiDS dataset \citep[collectively referred to as KiDZ;][]{vandenBusch20,vanDenBusch22}. These spectroscopic redshifts come from the Galaxy and Mass Assembly survey data release 4 \citep[GAMA-DR4;][]{Driver11,Liske15,Driver22}, the Sloan Digital Sky Survey \citep[SDSS;][]{Alam15}, and the 2-degree Field Lensing Survey \citep[2dFLens;][]{Blake16}. Each AMICO cluster has been assigned a bias-corrected redshift, with a resulting precision of $\sigma_z/(1+z) = 0.014$. The left panel of Fig.\ \ref{fig:redshift_distributions} displays the comparison of measured and de-biased AMICO cluster redshift distributions. \\
\indent As in \citet{Kuijken19}, \citetalias{Maturi25} employed the \texttt{Bayesian Photometric Redshift} \citep[][]{Benitez00} code to estimate galaxy photometric redshifts for cluster detection. The prior on redshift probabilities used in \citet{Kuijken19} was optimised for the weak-lensing analysis, reducing uncertainties and catastrophic outliers for faint, high-redshift galaxies. However, it introduced a significant bias for bright, low-redshift galaxies, which are important for cluster detection. For this reason, \citetalias{Maturi25} adopted instead the redshift prior used in the KiDS-DR3 analysis \citep{deJong17}. This choice results in a galaxy photometric redshift (photo-$z$) uncertainty of $\sigma_{z,{\rm g}}/(1+z) = 0.074$ for galaxies with $r < 23$, consistent with the findings of \citet{Kuijken19}. In the present work, we used photometric redshifts derived following this approach. \\
\indent To measure the weak-lensing profiles of AMICO galaxy clusters, we utilised the KiDS-1000 gold shear catalogue \citep{Wright20,Hildebrandt21,Giblin21}. This dataset contains approximately 21 million galaxies across an effective area of $777$ deg$^2$, with a weighted effective number density of $n^{\rm gold}_{\rm eff} = 6.17$ arcmin$^{-2}$. Shape measurements are based on the deep KiDS $r$-band imaging, and shear estimates are derived by employing the \textit{lensfit} algorithm \citep{Miller07,Fenech17}. Following \citet{Hildebrandt21}, we considered galaxies with redshifts in the range $z_{\rm g}\in[0.1,1.2]$. We used the same galaxy catalogue to measure the cluster--galaxy correlation function (Sect.\ \ref{sec:measure:2pt}). To ensure spatial matching, we restricted the cluster sample to the footprint of the galaxy sample, since the latter has a smaller effective area. For correlation function measurements, we also applied the $r<24$ magnitude cut, corresponding to the depth of the shallowest survey tile, to ensure homogeneity of the angular galaxy distribution \citepalias[following][]{Maturi25}. \\
\indent Following \citet{Hildebrandt21}, we reconstructed galaxy redshift distributions using self-organising maps (SOMs; \citealt{Kohonen1982,Wright20}). We did not use the galaxy redshift distributions from \citet{Hildebrandt21}, as our analysis adopts a different photo-$z$ estimation method and a different binning strategy. Instead, we applied SOMs to the spectroscopic sample from \citet{vanDenBusch22} and extended by \citet{Wright24}, comprising 93\,224 galaxies. While \citetalias{Maturi25} used a different method to estimate the unbiased cluster redshift distribution compared to the SOM-based approach we applied to galaxies, we do not expect this to introduce systematics. Indeed, as discussed in the following, our analysis accounts for the statistical uncertainties of both the galaxy and cluster redshift distributions. We employed the SOM algorithm introduced by \citet{Wright20} and subsequently implemented by \citet{Hildebrandt21} for KiDS-1000 redshift distribution calibration, based on the \texttt{kohonen} code \citep{Wehrens07,Wehrens18}. Following \citet{Wright20} and \citetalias{Lesci25}, we provided the SOM algorithm with photometric data from all 9 KiDS and VIKING bands, including their full colour space combinations. The SOM architecture employs a toroidal topology with a 50$\times$50 hexagonal grid. The middle panel in Fig.\ \ref{fig:redshift_distributions} shows the observed distribution of the full galaxy sample as a function of galaxy redshift, $z_{\rm g}$, along with the corresponding SOM-reconstructed distribution. The latter was employed in the cluster--galaxy correlation function modelling. The right panel of Fig.\ \ref{fig:redshift_distributions} displays three representative SOM-reconstructed background galaxy redshift distributions, each conditioned on a different cluster redshift. These distributions are measured in cluster redshift bins having a width of $\delta z=0.05$ and spanning the full sample range, based on a combination of colour and photo-$z$ background selections \citepalias[for details, we refer to][]{Lesci25}. We used these reconstructed background redshift distributions in the weak-lensing modelling.

\section{Measurements}\label{sec:measure}
In this section we present the measurements of stacked cluster lensing (Sect.\ \ref{sec:measure:wl}) and cluster--galaxy correlation function (Sect.\ \ref{sec:measure:2pt}). The lensing measurement pipeline follows that of \citetalias{Lesci25}, who carried out the mass calibration of the cluster sample analysed in this work. Null tests, including measurements of the cross-components of the lensing signal and around random points, are presented in \citetalias{Lesci25}. As discussed in Sect.\ \ref{sec:modelling:likelihood}, the statistical uncertainties related to multiplicative shear bias, galaxy redshift distributions, cluster orientation, and projection effects are propagated into the final covariance matrices. The impact of boost factors on our measurements, caused by projection effects, is discussed in Sect.\ \ref{sec:conclusions}. \\
\indent The properties of the cluster subsamples are reported in Table \ref{tab:sample}. As shown in Fig.\ 2 of \citetalias{Lesci25}, these cluster subsamples have a purity ranging from 97\% to 100\%. Consistent with \citetalias{Lesci25}, for all profile measurements we adopted a flat $\Lambda$CDM cosmology with a Hubble constant of $H_0 = 70~\mathrm{km~s^{-1}~Mpc^{-1}}$ and $\Omega_{\mathrm{m}} = 0.3$. Throughout this study, we considered different $\Omega_{\rm m}$ values to model our measurements. For the mass probe measurements considered in this work, cosmological dependence arises solely from the conversion to cluster-centric physical distances. As discussed in Sect.\ \ref{sec:modelling:stack}, we modelled the geometric distortions introduced by these assumptions.

\begin{table*}[t]
\small
\caption{\label{tab:sample}Properties of the cluster subsamples and main results.}
  \centering
    \begin{tabular}{c c c c c c c c c c c c} 
      $z_{\rm ob}$ range & $\lambda^*_{\rm ob}$ range & $N$  & Median $z_{\rm ob}$ & Median $\lambda^*_{\rm ob}$ & $\langle M_{200\rm m}^{g_{\rm t}}\rangle$ & $\langle r_{\rm sp}^{g_{\rm t}}\rangle$ & $\langle r_{\rm sp}^{w_{\rm cg}}\rangle$ & $\langle \mathcal{R}_{\rm sp}^{g_{\rm t}}\rangle$ & $\langle \mathcal{R}_{\rm sp}^{w_{\rm cg}}\rangle$ & $\langle \Gamma^{g_{\rm t}}\rangle$ & $\langle \Gamma^{w_{\rm cg}}\rangle$ \\
      \hline
      \rule{0pt}{4ex}
      $[0.10, 0.30)$ & $[20, 40)$ & 1411 & 0.24 & 25 & $0.70_{-0.06}^{+0.07}$ & $0.98_{-0.13}^{+0.13}$ & $0.88_{-0.07}^{+0.09}$ & $1.08_{-0.14}^{+0.15}$ & $0.97_{-0.09}^{+0.09}$ & $2.08_{-0.90}^{+1.11}$ & $2.97_{-0.77}^{+0.97}$ \\\rule{0pt}{2.5ex}
      $[0.10, 0.30)$ & $[40, 65)$ & 255 & 0.25 & 46 & $1.96_{-0.12}^{+0.11}$ & $1.36_{-0.18}^{+0.20}$ & $1.22_{-0.09}^{+0.12}$ & $1.06_{-0.13}^{+0.15}$ & $0.96_{-0.09}^{+0.09}$ & $2.20_{-0.93}^{+1.16}$ & $3.08_{-0.79}^{+0.98}$ \\\rule{0pt}{2.5ex}
      $[0.10, 0.30)$ & $[65, 210]$ & 47 & 0.25 & 78 & $4.63_{-0.37}^{+0.36}$ & $1.79_{-0.25}^{+0.28}$ & $1.59_{-0.12}^{+0.16}$ & $1.04_{-0.13}^{+0.15}$ & $0.94_{-0.09}^{+0.09}$ & $2.34_{-0.97}^{+1.22}$ & $3.27_{-0.83}^{+1.04}$ \\\rule{0pt}{3.5ex}
      $[0.30, 0.45)$ & $[25, 40)$ & 1536 & 0.38 & 30 & $0.83_{-0.06}^{+0.07}$ & $0.93_{-0.12}^{+0.14}$ & $0.84_{-0.06}^{+0.08}$ & $1.08_{-0.14}^{+0.15}$ & $0.99_{-0.09}^{+0.10}$ & $2.30_{-0.95}^{+1.19}$ & $3.11_{-0.79}^{+0.99}$ \\\rule{0pt}{2.5ex}
      $[0.30, 0.45)$ & $[40, 65)$ & 525 & 0.38 & 46 & $1.82_{-0.10}^{+0.10}$ & $1.19_{-0.16}^{+0.18}$ & $1.08_{-0.08}^{+0.10}$ & $1.07_{-0.14}^{+0.16}$ & $0.97_{-0.09}^{+0.10}$ & $2.41_{-0.99}^{+1.23}$ & $3.22_{-0.83}^{+1.04}$ \\\rule{0pt}{2.5ex}
      $[0.30, 0.45)$ & $[65, 210]$ & 90 & 0.38 & 76 & $4.23_{-0.31}^{+0.32}$ & $1.54_{-0.21}^{+0.24}$ & $1.39_{-0.11}^{+0.14}$ & $1.05_{-0.14}^{+0.16}$ & $0.95_{-0.09}^{+0.10}$ & $2.57_{-1.03}^{+1.30}$ & $3.44_{-0.86}^{+1.13}$ \\\rule{0pt}{3.5ex}
      $[0.45, 0.80]$ & $[30, 45)$ & 3616 & 0.64 & 35 & $0.98_{-0.08}^{+0.09}$ & $0.83_{-0.11}^{+0.13}$ & $0.77_{-0.06}^{+0.07}$ & $1.08_{-0.14}^{+0.16}$ & $0.99_{-0.10}^{+0.10}$ & $2.75_{-1.05}^{+1.39}$ & $3.55_{-0.89}^{+1.18}$ \\\rule{0pt}{2.5ex}
      $[0.45, 0.80]$ & $[45, 60)$ & 930 & 0.63 & 50 & $1.76_{-0.14}^{+0.13}$ & $1.00_{-0.14}^{+0.15}$ & $0.92_{-0.07}^{+0.09}$ & $1.07_{-0.14}^{+0.16}$ & $0.98_{-0.10}^{+0.10}$ & $2.84_{-1.09}^{+1.43}$ & $3.66_{-0.91}^{+1.24}$ \\\rule{0pt}{2.5ex}
      $[0.45, 0.80]$ & $[60, 210]$ & 320 & 0.61 & 68 & $3.22_{-0.27}^{+0.27}$ & $1.20_{-0.17}^{+0.19}$ & $1.10_{-0.09}^{+0.11}$ & $1.05_{-0.14}^{+0.16}$ & $0.96_{-0.09}^{+0.10}$ & $2.93_{-1.12}^{+1.49}$ & $3.82_{-0.96}^{+1.30}$ \\\rule{0pt}{2.5ex}
    \end{tabular}
    \tablefoot{\textit{Columns 1 and 2}: Bins of observed cluster photo-$z$, $z_{\rm ob}$, and intrinsic richness, $\lambda^*_{\rm ob}$. \textit{Column 3}: Number of clusters used for the measurements. \textit{Columns 4 and 5}: Median cluster redshift and richness. \textit{Column 6}: Posteriors on the mean mass, $\langle M_{200\rm m}^{g_{\rm t}}\rangle$ (in units of 
    $10^{14}h^{-1}{\rm M}_\odot$). \textit{Columns 7 and 8}: Splashback radius, $\langle r_{\rm sp}\rangle$ (in $h^{-1}$Mpc). \textit{Columns 9 and 10}: Normalised splashback radius, $\langle \mathcal{R}_{\rm sp}\rangle\equiv \langle r_{\rm sp}/r_{200\rm m}\rangle$. \textit{Columns 11 and 12}: Mass accretion rate, $\langle \Gamma\rangle$, which is dimensionless (see Eq.\ \ref{eq:Gamma}).
    The $g_{\rm t}$ and $w_{\rm cg}$ superscripts represent the results from weak lensing and the
    cluster--galaxy correlation function, respectively.}
\end{table*}

\subsection{Stacked weak-lensing profiles}\label{sec:measure:wl}
The tangential shear, $\gamma_{\rm t}$, is related to the excess surface density, $\Delta\Sigma_{\rm t}$, through the following relation \citep[see e.g.][]{Sheldon04}:
\begin{equation}
\Delta\Sigma_{\rm t}(R)=\overline{\Sigma}(<R)-\Sigma(R)=\Sigma_{\rm crit}\gamma_{\rm t}(R),
\end{equation}
where $\Sigma(R)$ denotes the surface mass density at radius $R$, $\overline{\Sigma}(<R)$ is its mean enclosed within the radius $R$, and $\Sigma_{\rm crit}$ is the critical surface density, which is expressed as \citep{Bartelmann01}
\begin{equation}\label{eq:sigma_crit}
\Sigma_{\rm crit} \equiv \frac{c^2}{4\pi G}\frac{D_{\rm s}}{D_{\rm l}D_{\rm ls}}\,,
\end{equation}
where $c$ is the speed of light, $G$ is Newton's gravitational constant, while $D_{\rm s}$, $D_{\rm l}$, and $D_{\rm ls}$ are the observer-source, observer-lens, and lens-source angular diameter distances, respectively.\\
\indent The weak-lensing observable linked to galaxy ellipticities is the reduced shear, namely $g=\gamma/(1-\kappa)$, where $\kappa\equiv\Sigma/\Sigma_{\rm crit}$ is the convergence \citep{Schneider1995}. We estimated the reduced tangential shear profile, $g_{\rm t}$, of a cluster labelled as $k$ as \citep[see e.g.][]{Viola15}
\begin{equation}\label{eq:DeltaSigma_measure}
g_{\rm t,k}(R_j)=\left( \frac{\sum_{i\in j}\;w_i\,e_{\rm t,i}}{\sum_{i\in j}\;w_i} \right)\frac{1}{1+\mathcal{M}_j},
\end{equation}
where $e_{\rm t,i}$ is the tangential component of the ellipticity of the $i$th background galaxy, $j$ is the radial annulus index, with an associated average projected radius $R_j$, corresponding to the central value of the radial bin. The impact of this radial point definition on our results is expected to be negligible. Indeed, \citetalias{Lesci25} demonstrated that for our sample, alternative estimates of the effective radius differ from $R_j$ by only 0.1\%. In addition, $w_i$ is the statistical weight assigned to the measure of the source ellipticity of the $i$th background galaxy \citep{Sheldon04}, while $\mathcal{M}_j$ denotes the average multiplicative shear bias, defined as
\begin{equation}
\mathcal{M}_j=\frac{\sum_{i\in j}\;w_i\,m_i}{\sum_{i\in j}\;w_i},
\end{equation}
where $m_i$ is the multiplicative shear bias of the $i$th background galaxy. Following \citetalias{Lesci25}, for each background source we drew a value of $m$ from a uniform distribution in the range $[-2\sigma_m, 2\sigma_m]$, where $\sigma_m=0.02$ corresponds to the largest 1$\sigma$ interval of the $m$ distributions obtained by \citet{Giblin21}. We used this approximation because the $m$ estimates by \citet{Giblin21} were derived from redshift distributions that do not correspond to those considered in this work. The statistical uncertainty on $m$, $\sigma_m$, is propagated into the final results by including it in the covariance matrix (see Sect.\ \ref{sec:modelling:likelihood}). We neglected the shear additive bias term, because \citet{Giblin21} constrained it to be $10^{-4}$, which is an order of magnitude smaller than our stacked $g_{\rm t}$ uncertainties. The background galaxy samples are defined through the combination of the photo-$z$ and colour selections defined in \citetalias{Lesci25}. In the modelling described in Sect.\ \ref{sec:modelling}, we used the background redshift distributions reconstructed through SOMs, as described in Sect.\ \ref{sec:dataset}. \\
\indent The stacked reduced shear profile in the $p$th bin of observed intrinsic richness, $\Delta\lambda^*_{{\rm ob},p}$, and in the $q$th bin of observed cluster redshift, $\Delta z_{{\rm ob},q}$, is expressed as
\begin{equation}\label{eq:DeltaSigma_measure_stack}
g_{\rm t}(R_j,\Delta\lambda^*_{{\rm ob},p},\Delta z_{{\rm ob},q}) = \frac{\sum_{k\in \Delta\lambda^*_{{\rm ob},p},\Delta z_{{\rm ob},q}}W_{k,j}\,g_{\rm t,k}(R_j)}{\sum_{k\in \Delta\lambda^*_{{\rm ob},p},\Delta z_{{\rm ob},q}}W_{k,j}},
\end{equation}
where $g_{\rm t,k}$ is given by Eq.\ \eqref{eq:DeltaSigma_measure}, $k$ runs over all clusters falling in the bins of $\lambda^*$ and $z$, while $W_{k,j}$ is the total weight
for the $j$th radial bin of the $k$th cluster, estimated as
\begin{equation}\label{eq:W_lensing}
W_{k,j} = \sum_{i\in j} w_i\,,
\end{equation}
where $i$ runs over the background galaxies in the $j$th radial bin. We used the same $\lambda^*$ and $z$ bins as \citetalias{Lesci25}, shown in Table \ref{tab:sample}. This $\lambda^*$ and $z$ selection yields a sample of 9049 clusters, which reduce to 8730 due to the smaller area covered by the shear sample compared to the cluster sample (see Sect.\ \ref{sec:dataset}). We adopted the same minimum cluster-centric projected distance of 400 $h^{-1}$kpc as \citetalias{Lesci25}, but extended the maximum radius to 5 $h^{-1}$Mpc in order to include measurements of the infalling region of galaxy clusters. Ten logarithmically spaced radial bins are used. \\
\indent The bootstrap-estimated covariance matrix for each stacked $g_{\rm t}(R)$ measurement derives from 10\,000 re-samplings of the cluster profiles within the stack. This covariance estimate incorporates the intrinsic scatter in the observable-mass and concentration-mass relations, along with miscentring effects and the contribution from the large-scale structure, though it does not account for the covariance due to background galaxies shared between clusters. However, as demonstrated by \citet{McClintock19}, for Stage-III surveys such shared-source contributions remain subdominant to shape noise and large-scale structure effects. Finally, we neglected the covariance across different redshift and richness bins, as its impact has been shown to be negligible in Stage-III surveys \citep{McClintock19}.

\subsection{Cluster--galaxy correlation function}\label{sec:measure:2pt}
The projected cluster--galaxy correlation function, denoted as $w_{\rm cg}$, in the $p$th bin of intrinsic richness, $\Delta\lambda^*_{{\rm ob},p}$, and in the $q$th bin of cluster redshift, $\Delta z_{{\rm ob},q}$, is measured using the Landy-Szalay estimator \citep{Landy1993}, expressed as follows:
\begin{equation}\label{eq:LS_estimator}
w_{\rm cg}(R_j,\Delta\lambda^*_{{\rm ob},p},\Delta z_{{\rm ob},q}) = \frac{ \mathcal{D}_{\rm c}\mathcal{D}_{\rm g}(R_j) - \mathcal{D}_{\rm c}\mathcal{R}_{\rm g}(R_j) - \mathcal{R}_{\rm c}\mathcal{D}_{\rm g}(R_j) }{ \mathcal{R}_{\rm c}\mathcal{R}_{\rm g}(R_j) } +1\,.
\end{equation}
Compared to the Davis-Peebles estimator \citep{Davis1983}, Eq.\ \eqref{eq:LS_estimator} reduces variance and edge effects. In Eq.\ \eqref{eq:LS_estimator}, $R_j$ is the central point of the $j$th radial annulus, while $\mathcal{D}_{\rm c}\mathcal{D}_{\rm g}(R_j)$, $\mathcal{D}_{\rm c}\mathcal{R}_{\rm g}(R_j)$, $\mathcal{R}_{\rm c}\mathcal{D}_{\rm g}(R_j)$, and $\mathcal{R}_{\rm c}\mathcal{R}_{\rm g}(R_j)$ are the normalised data cluster--data galaxy, data cluster--random galaxy, random cluster--data galaxy, random cluster--random galaxy pairs, respectively. In Eq.\ \eqref{eq:LS_estimator}, we adopted the same $R_j$ points used for weak-lensing measurements. \\
\indent To measure $w_{\rm cg}$, we used the complete KiDS-1000 gold shear catalogue, imposing the magnitude cut $r<24$ (see Sect.\ \ref{sec:dataset}). As this cut corresponds to the depth of the shallowest tile, it ensures a homogeneous angular galaxy distribution. Furthermore, in Eq.\ \eqref{eq:LS_estimator}, for each cluster we include all galaxies at a projected separation $R_j$ along the line of sight. As a result, the $w_{\rm cg}$ estimator is independent of cluster membership modelling. The $w_{\rm cg}$ model is defined consistently with this approach (see Sect.\ \ref{sec:modelling:single_probes:w}). \\
\indent We constructed random cluster and random galaxy samples 80 times larger than the observed ones. The extraction of random (RA, Dec) pairs accounts for the KiDS survey angular mask. We assigned random cluster redshifts by reshuffling the observed mean redshifts, which were subsequently used to define cluster-centric radial bins. For the random galaxy sample, no redshift information is required since galaxy redshifts do not enter the $w_{\rm cg}$ estimator (Eq.\ \ref{eq:LS_estimator}). The adopted cluster $\lambda^*$, $z$, and $R$ bins are those used for the weak-lensing measurements presented in Sect.\ \ref{sec:measure:wl}. Jackknife resampling was used to estimate the covariance matrix, with KiDS tiles defining the resampling regions, which cover an area of about 1 deg$^2$. As in Sect.\ \ref{sec:measure:wl}, we neglected the correlation between $\lambda^*$ and $z$ bins.

\section{Modelling}\label{sec:modelling}
In this section we present the likelihood framework used to constrain cluster splashback radii. In Sect.\ \ref{sec:modelling:single} we describe the adopted 3D halo mass profile. Sections \ref{sec:modelling:single_probes:g} and \ref{sec:modelling:single_probes:w} introduce the models for $g_{\rm t}$ and $w_{\rm wg}$ for individual haloes, obtained by projecting the 3D profile along the line of sight. In Sect.\ \ref{sec:modelling:stack} we describe the modelling of the stacked $g_{\rm t}$ and $w_{\rm wg}$ signals, explicitly accounting for the scatter in the observable-mass relation, halo mass function, and selection effects. Section \ref{sec:modelling:splashback} describes the splashback radius estimator, while in Sect.\ \ref{sec:modelling:likelihood} we introduce the likelihood function, covariance matrices, and parameter priors.

\subsection{3D model for individual haloes}\label{sec:modelling:single}
This section presents the theoretical description of individual 3D mass profiles, characterised by the \citet[][DK14]{Diemer14} profile, widely used in the modelling of observational data \citep{Baxter17,Rana23,Giocoli24,Joshi25}. The corresponding excess 3D density profile is expressed as follows:
\begin{equation}\label{eq:DK14}
\Delta\rho(r) = \rho_{\rm s} \, \exp\bigg\{-\frac{2}{\alpha}\,\bigg[\bigg(\frac{r}{r_{\rm s}}\bigg)^\alpha -1\bigg]\bigg\} \, \,f_{\rm trans} +\rho_{\rm outer}\,,
\end{equation}
where
\begin{equation}\label{eq:DK14:transition}
f_{\rm trans} = \bigg[1 + \bigg(\frac{r}{r_{\rm t}}\bigg)^\beta\bigg]^{-{\frac{\gamma}{\beta}}}\,,
\end{equation}
\begin{equation}\label{eq:DK14:infall}
\rho_{\rm outer} = \rho_{\rm m}\,b_{\rm e} \,\bigg(\dfrac{r}{5r_{\rm 200m}}\bigg)^{-s_{\rm e}}\,.
\end{equation}
Equation \eqref{eq:DK14} is an \citet{Einasto1965} profile multiplied by the transition factor, $f_{\rm trans}$. Here, the dependence of $\Delta\rho$ on $M_{200\rm m}$ and $z$ is not explicitly shown, for brevity. In Eq.\ \eqref{eq:DK14}, $\rho_{\rm s}$ is the characteristic density and $r_{\rm s}$ is the scale radius, expressed as $r_{\rm s} = r_{200\rm m}/c_{200\rm m}$, where $c_{200\rm m}$ is the halo concentration. As detailed in Sect.\ \ref{sec:modelling:stack}, we modelled $c_{\rm 200m}$ via a log-linear scaling relation with $M_{\rm 200m}$ and redshift, with the amplitude empirically constrained by our data. For the inner slope of the Einasto profile, we assumed $\alpha=0.155 + 0.0095 \nu_{\rm vir}^2$ \citep{Gao08}. Here, $\nu^2_{\rm vir}=\delta^2_{\rm c}(z)/\sigma^2(M_{\rm vir})$ represents the peak height of the virial halo mass, namely $M_{\rm vir}$, while $\delta_{\rm c}(z)$ is the linear theory critical overdensity required for spherical collapse divided by the growth factor, and $\sigma^2(M_{\rm vir})$ is the mass variance. For the virial overdensity definition, we adopted the fitting function by \citet{Bryan1998}. To be consistent with the $\alpha$ definition by \citet{Gao08}, we converted $M_{\rm 200m}$ into $M_{\rm vir}$ assuming a \citet*[][NFW]{NFW} profile. Furthermore, we note that by treating $c_{\rm 200m}$ as a free parameter, any possible systematic errors arising from the assumptions on $\alpha$ are mitigated. \\
\indent The Einasto profile in Eq.\ \eqref{eq:DK14} is multiplied by a transition factor, $f_{\rm trans}$ (defined in Eq.\ \ref{eq:DK14:transition}), which smoothly connects the inner halo component to the external large-scale term, namely $\rho_{\rm outer}$ (defined in Eq.\ \ref{eq:DK14:infall}). The transition factor depends on: (i) the truncation radius, $r_{\rm t}=F_{\rm t}r_{200\rm m}$, where $F_{\rm t}$ is the truncation factor, (ii) $\beta$, which defines the sharpness of the transition from the one-halo to the outer profile, and (iii) $\gamma$, which describes the steepness of the profile at $r\sim r_{200\rm m}$. We expressed $\gamma$ as $\gamma = \gamma_0 \nu_{\rm vir}$, following \citetalias{Diemer14}. Furthermore, as discussed in Sect.\ \ref{sec:modelling:likelihood}, we assumed Gaussian priors on $F_{\rm t}$, $\beta$, and $\gamma_0$.\\
\indent Equation \eqref{eq:DK14:infall} models the matter density distribution at $r>r_{200\rm m}$, and it is expressed as a power law that multiplies the mean background density of the Universe, $\rho_{\rm m}$. Following the \citetalias{Diemer14} prescription, we defined the pivot radius as $5$ times $r_{\rm 200m}$. The parameters $b_{\rm e}$ and $s_{\rm e}$ characterise the amplitude and slope of the outer profile, respectively, and are treated as free parameters in our analysis. In simulations, both the bias parameter, $b_{\rm e}$, and the slope, $s_{\rm e}$, exhibit a mild dependence on the peak height \citep{Diemer14}. While $s_{\rm e}$ shows only a weak redshift evolution, this dependence is somewhat stronger for $b_{\rm e}$. Thus, in our modelling, we allowed $b_{\rm e}$ to vary as a function of redshift, assuming the following expression: 
\begin{equation}\label{eq:b_e}
b_{\rm e}=b_{\rm e,0}\left[\frac{1+z}{1+z_{\rm piv}}\right]^{b_{{\rm e},z}}\,,
\end{equation}
where $z_{\rm piv}=0.4$. We note that, differently from the \citetalias{Diemer14} functional form, Eq.\ \eqref{eq:DK14:infall} does not include $\rho_{\rm m}$ as an additive term. In fact, the mass profile probes considered in this study, namely weak lensing and  the correlation function, measure the excess of mass compared to the background. On the other hand, as detailed in Sect.\ \ref{sec:modelling:splashback}, the $\rho_{\rm m}$ contribution is considered in the computation of the splashback radius. The six free parameters describing the halo profiles ($F_t$, $\beta$, $\gamma_0$, $b_{{\rm e},0}$, $b_{{\rm e},z}$, and $s_{\rm e}$) are summarised in the top part of Table \ref{tab:priors_and_posteriors}. \\

\subsection{$g_{\rm t}$ model for individual haloes}\label{sec:modelling:single_probes:g}
The tangential reduced shear component of a halo, $g_{\rm t}$, is expressed as \citep{Seitz97}
\begin{equation}\label{eq:g}
g_{\rm t}(R,M,z) = \frac{\Delta\Sigma_{\rm t}(R,M,z)\,\langle\Sigma_{\rm crit}^{-1}(z)\rangle}{1 - \Sigma(R,M,z)\,\langle\Sigma_{\rm crit}^{-1}(z)\rangle^{-1}\,\langle\Sigma_{\rm crit}^{-2}(z)\rangle}\,,
\end{equation}
where $R$ is the projected radius, $\Sigma$ is the surface mass density, obtained by integrating Eq.\ \eqref{eq:DK14} as
\begin{equation}
\Sigma(R,M,z) = 2\int\limits_0^{R_{\rm max}} {\rm d} \chi\,\Delta\rho\left(\sqrt{R^2+\chi^2},M,z\right)\,,
\end{equation}
where we set $R_{\rm max}=40$ $h^{-1}$Mpc \citep[following][]{More16,Baxter17,Shin19}. In Eq.\ \eqref{eq:g}, $\Delta\Sigma_{\rm t}$ is the excess surface mass density, having the expression
\begin{equation}\label{eq:DeltaSigma_cen}
\Delta\Sigma_{\rm t}(R) = \frac{2}{R^2}\int\limits_0^R \mathrm{d} r\,\, r\, \Sigma(r) -\Sigma(R)\,.
\end{equation}
In Eq.\ \eqref{eq:g}, $\langle\Sigma_{\rm crit}^{-\eta}\rangle$ is defined as
\begin{equation}
\langle\Sigma_{\rm crit}^{-\eta}(z)\rangle = \int\limits_{z_{\rm g}>z}{\rm d}z_{\rm g}\, \Sigma_{\rm crit}^{-\eta}(z_{\rm g},z)\,n(z_{\rm g}\,|\,z)\,,
\end{equation}
where $\eta=1,2$, $\Sigma_{\rm crit}$ is given by Eq.\ \eqref{eq:sigma_crit}, $z_{\rm g}$ is the galaxy redshift, while $n(z_{\rm g}\,|\,z)$ is the normalised true background redshift distribution given a cluster redshift $z$, reconstructed through SOMs (Sect.\ \ref{sec:dataset}). 
We remark that Eq.\ \eqref{eq:g} describes the reduced shear profiles of perfectly centred cluster detections. The total reduced shear includes the contribution by miscentred clusters and is expressed as follows:
\begin{equation}\label{eq:g_tot}
g_{\rm t,\rm tot} = (1-f_{\rm off})\, g_{\rm t} + f_{\rm off}\, g_{\rm t,\rm off}\,,
\end{equation}
where $f_{\rm off}$ is the fraction of miscentred clusters, $g_{\rm t}$ is given by Eq.\ \eqref{eq:g}, while $g_{\rm t,\rm off}$ is the miscentred tangential reduced shear, whose derivation is detailed in Appendix \ref{app:g_off}. 

\subsection{$w_{\rm cg}$ model for individual haloes}\label{sec:modelling:single_probes:w}
We express the 2D cluster--galaxy correlation function as follows (see Appendix \ref{app:w_model} for its derivation):
\begin{equation}\label{eq:w}
w_{\rm cg}(R,M,z) = \frac{\langle b_{\rm g}(M,z)\rangle}{\rho_{\rm m}(z)}\int\limits_0^\infty \! \mathrm{d}z_{\rm g} \,\Delta\rho\left[f(R,z_{\rm g},z),M,z\right]\,n(z_{\rm g})\,,
\end{equation}
where $\Delta\rho$ is the \citetalias{Diemer14} profile (Eq.\ \ref{eq:DK14}), while $n(z_{\rm g})$ is the normalised galaxy redshift distribution, reconstructed through SOMs and shown in the middle panel of Fig.\ \ref{fig:redshift_distributions} (see also Sect.\ \ref{sec:dataset}). For the integration in Eq.\ \eqref{eq:w}, we interpolated the discretised integrand. In Eq.\ \eqref{eq:w}, $f(R,z_{\rm g},z)$ is the physical distance between a galaxy at $z_{\rm g}$ and a cluster at $z$, given a projected cluster-centric distance $R$, expressed as follows:
\begin{equation}
f(R,z_{\rm g},z) = \sqrt{R^2+\left[\frac{\chi(z_{\rm g})-\chi(z)}{1+z}\right]^2}\,,
\end{equation}
where $\chi$ is the comoving distance. Furthermore, $b_{\rm g}$ in Eq.\ \eqref{eq:w} is the galaxy bias, treated as independent of $R$ (see Appendix \ref{app:w_model}). As discussed in Sects.\ \ref{sec:modelling:stack} and \ref{sec:modelling:likelihood}, we did not model the $b_{\rm g}$ dependence on the true halo mass and redshift. Instead, for each stack of clusters defined in bins of observed intrinsic richness $\lambda^*_{\rm ob}$ and cluster redshift $z_{\rm ob}$, we introduced the average bias $\langle b_{\rm g}(\Delta\lambda^*_{\rm ob}, \Delta z_{\rm ob})\rangle$ as a free parameter. We did not model miscentring in Eq. \eqref{eq:w} because it primarily affects mass estimates. Since we adopted the mass posteriors from the $g_{\rm t}$ analysis as priors for the $w_{\rm cg}$ modelling (see Sect. \ref{sec:modelling:likelihood}), this effect is inherently accounted for in our approach.

\subsection{Expected values for stacks of clusters}\label{sec:modelling:stack}
\begin{table*}[t]
\caption{\label{tab:priors_and_posteriors}Free base parameters considered in the analysis. }
  \centering
    \begin{tabular}{c c c c c c c c} 
      Parameter & Description & $g_{\rm t}$ prior & $w_{\rm cg}$ prior & $g_{\rm t}$ posterior & $w_{\rm cg}$ posterior \\ 
      \hline
      \rule{0pt}{4ex}
      $F_{\rm t}$ & Truncation factor of the one-halo profile & $\mathcal{N}(1.495, 0.3)$ & $\mathcal{N}(1.495, 0.3)$ & $1.40^{+0.29}_{-0.22}$ & $1.51^{+0.25}_{-0.21}$ \\ \rule{0pt}{2.5ex}
      $\beta$ & Sharpness of the transition from one-halo to two-halo & $\mathcal{N}(4, 1.6)$ & $\mathcal{N}(4, 1.6)$ & --- & $3.35^{+0.94}_{-1.01}$ \\ \rule{0pt}{2.5ex}
      $\gamma_0$ & Steepness of the profile at $r\sim r_{200\rm m}$ & $\mathcal{N}(4, 1.6)$ & $\mathcal{N}(4, 1.6)$ & --- & --- \\ \rule{0pt}{2.5ex}
      $b_{{\rm e},0}$ & Normalisation of the outer profile's amplitude & $[0.5,4]$ & $[0.5,4]$ & --- & $1.97^{+0.23}_{-0.19}$ \\ \rule{0pt}{2.5ex}\rule{0pt}{2.5ex}
      $b_{{\rm e},z}$ & Redshift evolution of the outer profile's amplitude & $[-1,1]$ & $[-1,1]$ & --- & --- \\ \rule{0pt}{2.5ex}
      $s_{\rm e}$ & Slope of the outer profile & $[0.5,2]$ & $[0.5,2]$ & --- & $1.461^{+0.052}_{-0.046}$ \\ \rule{0pt}{2.5ex}
      $f_{\rm off}$ & Fraction of miscentred clusters & $\mathcal{N}(0.3, 0.1)$ & --- & --- & --- \\ \rule{0pt}{2.5ex}
      $\sigma_{{\rm off}}$ & Miscentring scale (in $h^{-1}$Mpc) & [0, 0.5] & --- & $0.21^{+0.11}_{-0.11}$ & --- \\ \rule{0pt}{2.5ex}
      $A$ & Amplitude of the $\log\lambda^*-\log M_{200\rm m}$ relation & [-2, 2] & $g_{\rm t}$ posterior & $-0.27^{+0.03}_{-0.05}$ & --- \\ \rule{0pt}{2.5ex}
      $B$ & Slope of the $\log\lambda^*-\log M_{200\rm m}$ relation & [0, 3] & $g_{\rm t}$ posterior  & $0.59^{+0.06}_{-0.05}$ & --- \\ \rule{0pt}{2.5ex}
      $C$ & Redshift evolution of the $\log\lambda^*-\log M_{200\rm m}$ relation & [-3, 3] & $g_{\rm t}$ posterior & $0.28^{+0.29}_{-0.30}$ & --- \\ \rule{0pt}{2.5ex}
      $\sigma_{\rm intr}$ & Intrinsic scatter of the $\log\lambda^*-\log M_{200\rm m}$ relation & [0.01, 0.5] & $g_{\rm t}$ posterior & $0.07^{+0.04}_{-0.03}$ & --- \\ \rule{0pt}{2.5ex}
      $\log c_0$ & Amplitude of the $\log c_{200\rm m}-\log M_{200\rm m}$ relation & $[0, 1.3]$ & $[0, 1.3]$ & $0.68^{+0.17}_{-0.19}$ & $0.82^{+0.19}_{-0.29}$ \\ \rule{0pt}{2.5ex}
      $(s,q)$ & Parameters of the mass function correction factor & $\mathcal{N}(\mu_{\rm \scriptscriptstyle HMF}, C_{\rm \scriptscriptstyle HMF})$ & $\mathcal{N}(\mu_{\rm \scriptscriptstyle HMF}, C_{\rm \scriptscriptstyle HMF})$ & --- & --- \\ \rule{0pt}{2.5ex}
      $\log b_{{\rm g},0}$ & Logarithmic amplitude of the galaxy bias function & --- & [-3, 1] & --- & $0.068^{+0.062}_{-0.067}$ \\ \rule{0pt}{2.5ex}
      $b_{{\rm g},\lambda^*}$ & $\lambda^*$ evolution of the galaxy bias function & --- & [-2, 2] & --- & $-0.10^{+0.11}_{-0.11}$ \\ \rule{0pt}{2.5ex}
      $b_{{\rm g},z}$ & $z$ evolution of the galaxy bias function & --- & [-2, 2] & --- & $0.31^{+0.35}_{-0.33}$ \\
    \end{tabular}
  \tablefoot{Derived parameters, such as $r_{\rm sp}$, are not reported. \textit{Columns 1 and 2}: Symbols and descriptions of the parameters, respectively. \textit{Columns 3 and 4}: Parameter priors adopted for the modelling of $g_{\rm t}$ and $w_{\rm cg}$, respectively. Here, a range represents a uniform prior, while $\mathcal{N}(\mu,\sigma)$ stands for a Gaussian prior with mean $\mu$ and standard deviation $\sigma$. \textit{Columns 5 and 6}: Median values of the 1D marginalised posteriors, along with the 16th and 84th percentiles. The posterior is not reported in cases where it closely aligns with the prior.}
\end{table*}
The expected value of the stacked reduced shear in a bin of observed intrinsic richness, $\Delta\lambda^*_{\rm ob}$, and redshift, $\Delta z_{\rm ob}$, is expressed as
\begin{align}\label{eq:g_stack}
\langle &g_{\rm t}(R^{\rm test},\Delta\lambda^*_{\rm ob},\Delta z_{\rm ob})\rangle =
\frac{\mathcal{P}_{\rm cl}(\Delta \lambda^*_{\rm ob},\Delta z_{\rm ob})\,\langle\mathcal{P}_{\rm bkg}(\Delta z_{\rm ob})\rangle}{\langle n(\Delta \lambda^*_{\rm ob},\Delta z_{\rm ob})\rangle}\,\times\nonumber\\
&\times\int\limits_0^\infty{\rm d}z_{\rm tr}\,\frac{{\rm d}^2 V}{{\rm d} z_{\rm tr}{\rm d}\Omega} \int\limits_0^\infty{\rm d}M\, g_{\rm t,\rm tot}(R^{\rm test},M,z_{\rm tr}) \, \mathcal{S}(M,z_{\rm tr}, \Delta\lambda^*_{\rm ob},\Delta z_{\rm ob})\,,
\end{align}
where $\mathcal{P}_{\rm cl}$ is the cluster sample purity, based on the Selection Function extrActor \citep[SinFoniA;][]{Maturi19,Maturi25} mock catalogues, while $\mathcal{P}_{\rm bkg}$ is the purity of the background galaxy sample, derived through SOMs. For details on these purity parameters, we refer to \citetalias{Lesci25}. In Eq.\ \eqref{eq:g_stack}, $z_{\rm tr}$ is the true redshift, $V$ is the co-moving volume, ${\rm d}\Omega$ is the solid angle element, $M$ is the true mass, while $g_{\rm t,\rm tot}$ is given by Eq.\ \eqref{eq:g_tot} and, in order to model geometric distortions, it is computed at the test radius, $R^{\rm test}$,
\begin{equation}\label{eq:R_test}
R^{\rm test} = \theta D^{\rm test}_{\rm l}  = R^{\rm fid} \, \frac{D^{\rm test}_{\rm l}}{D^{\rm fid}_{\rm l}}\,.
\end{equation}
Here, $\theta$ is the angular separation from the cluster centre, $R^{\rm fid}$ is the projected radius in the fiducial cosmology adopted for measurements (Sect.\ \ref{sec:measure}), while $D_{\rm l}^{\rm fid}$ and $D_{\rm l}^{\rm test}$ are the angular diameter distances in the fiducial and test cosmologies, respectively. The last factor in the integrand of Eq.\ \eqref{eq:g_stack} has the following expression:
\begin{align}\label{eq:selection_function}
\mathcal{S}(M&,z_{\rm tr}, \Delta\lambda^*_{\rm ob},\Delta z_{\rm ob}) = \frac{{\rm d} n(M,z_{\rm tr})}{{\rm d} M}\,\mathcal{B}_{\rm\scriptscriptstyle HMF}(M) \int\limits_{\Delta z_{\rm ob}}{\rm d} z_{\rm ob}\, P(z_{\rm ob}|z_{\rm tr}) \,\times\nonumber\\
&\times\int\limits_0^\infty{\rm d}\lambda^*_{\rm tr}\,\mathcal{C}_{\rm cl}(\lambda^*_{\rm tr},z_{\rm tr})\,P(\lambda^*_{\rm tr}|M,z_{\rm tr})\int\limits_{\Delta\lambda^*_{\rm ob}}{\rm d} \lambda^*_{\rm ob} \,\,P(\lambda^*_{\rm ob}|\lambda^*_{\rm tr},z_{\rm tr})\,,
\end{align}
where ${\rm d} n(M,z_{\rm tr})/{\rm d} M$ is the halo mass function, for which we adopted the model by \citet{Tinker08}, while $\mathcal{B}_{\rm\scriptscriptstyle HMF}(M)$ is the halo mass function bias. Following \citet{Costanzi19}, $\mathcal{B}_{\rm\scriptscriptstyle HMF}(M)$ is expressed as
\begin{equation}\label{eq:B_HMF}
\mathcal{B}_{\rm\scriptscriptstyle HMF}(M) = s\log \frac{M}{M^*}+q\,,
\end{equation}
where $M^*$ is expressed in $h^{-1}$M$_\odot$ and $\log M^*=13.8$, $q$ and $s$ are free parameters of the model (see Sect.\ \ref{sec:modelling:likelihood}). In Eq.\ \eqref{eq:selection_function}, $z_{\rm ob}$, $\lambda^*_{\rm ob}$, and $\lambda^*_{\rm tr}$ are the observed redshift, observed intrinsic richness, and true intrinsic richness, respectively. $P(z_{\rm ob}|z_{\rm tr})$ is a Gaussian probability density function with mean corresponding to $z_{\rm tr}$ and a standard deviation of $0.014(1+z_{\rm tr})$ (see Sect.\ \ref{sec:dataset}), while $P(\lambda^*_{\rm ob}|\lambda^*_{\rm tr},z_{\rm tr})$ is a Gaussian whose mean is expressed as 
\begin{equation}\label{eq:Plambda_mean_final}
    \mu_{\lambda^*} = \lambda^*_{\rm tr} + \lambda^*_{\rm tr} \exp[- \lambda^*_{\rm tr} \, (A_\mu + B_\mu \, z_{\rm tr})]\,,
\end{equation}
while its standard deviation has the form
\begin{equation}\label{eq:Plambda_std_final}
    \sigma_{\lambda^*} = A_\sigma \lambda^*_{\rm tr} \exp(- B_\sigma \lambda^*_{\rm tr})\,.
\end{equation}
In Eqs.\ \eqref{eq:Plambda_mean_final} and \eqref{eq:Plambda_std_final}, $A_\mu=0.198$, $B_\mu = -0.179$, $A_\sigma = 0.320$, and $B_\sigma = 0.011$. We refer to \citetalias{Lesci25} for details on the derivation of $P(\lambda^*_{\rm ob}|\lambda^*_{\rm tr},z_{\rm tr})$, its coefficients, and of the cluster sample completeness, $\mathcal{C}_{\rm cl}$, appearing in Eq.\ \eqref{eq:selection_function} and based on SinFoniA. As discussed in \citetalias{Lesci25}, $\sigma_x$ in Eq.\ \eqref{eq:Plambda_std_final} is up to $35\%$ larger than Poisson noise due to masking, blending of cluster detections, and projection effects. Furthermore, $\mu_x$ (Eq.\ \ref{eq:Plambda_mean_final}) deviates from $\lambda^*_{\rm tr}$ mainly due to projection effects. \\
\indent The probability density distribution $P(\lambda^*_{\rm tr}|M,z_{\rm tr})$ follows a log-normal form, with its mean determined by the $\log\lambda^*-\log M_{200\rm m}$ scaling relation and its dispersion set by the intrinsic scatter, $\sigma_{\rm intr}$:
\begin{align}\label{eq:scalingrelation_PDF}
P(\lambda^*_{\rm tr}|M,z_{\rm tr})=&\frac{1}{\ln(10)\lambda^*_{\rm tr}\sqrt{2\pi}\sigma_{\rm intr}}\exp\left(-\frac{[\log \lambda^*_{\rm tr} - \mu(M,z_{\rm tr})]^2}{2\sigma^2_{\rm intr}}\right)\,.
\end{align}
Here, $\mu(M,z_{\rm tr})$ is the mean of the distribution, expressed as
\begin{align}\label{eq:scalingrelation}
\mu(M,z_{\rm tr})=A+B\log\frac{M}{M_{\rm piv}}+C\log\frac{H(z_{\rm tr})}{H(z_{\rm piv})} + \log\lambda^*_{\rm piv}\,,
\end{align}
where $M_{\rm piv}=10^{14}h^{-1}M_\odot$, $z_{\rm piv}=0.4$, and $\lambda^*_{\rm piv}=50$ are the redshift, mass, and intrinsic richness pivots, respectively. The penultimate term in Eq.\ \eqref{eq:scalingrelation}, including the Hubble function $H(z)$, captures deviations from the predictions from the self-similar growth scenario \citep{Sereno15}. As described in Sect.\ \ref{sec:modelling:likelihood}, $A$, $B$, $C$, and $\sigma_{\rm intr}$ in Eqs.\ \eqref{eq:scalingrelation_PDF} and \eqref{eq:scalingrelation} are treated as free parameters in our modelling (see also Table \ref{tab:priors_and_posteriors}). Lastly, in Eq.\ \eqref{eq:g_stack}, $\langle n(\Delta \lambda^*_{\rm ob},\Delta z_{\rm ob})\rangle$ is the expected density of observed haloes and has the form
\begin{equation}
\langle n(\Delta \lambda^*_{\rm ob},\Delta z_{\rm ob})\rangle = \int\limits_0^\infty{\rm d}z_{\rm tr}\,\frac{{\rm d}^2 V}{{\rm d} z_{\rm tr}{\rm d}\Omega} \int\limits_0^\infty{\rm d}M\, \mathcal{S}(M,z_{\rm tr}, \Delta\lambda^*_{\rm ob},\Delta z_{\rm ob})\,.
\end{equation}
The expected cluster--galaxy correlation function in the bins $\Delta \lambda^*_{\rm ob}$ and $\Delta z_{\rm ob}$ is expressed as
\begin{align}\label{eq:w_stack}
\langle &w_{\rm cg}(R^{\rm test},\Delta\lambda^*_{\rm ob},\Delta z_{\rm ob})\rangle =
- \mathcal{I}(\Delta\lambda^*_{\rm ob},\Delta z_{\rm ob}) + \frac{\langle b_{\rm g}(\Delta \lambda^*_{\rm ob},\Delta z_{\rm ob})\rangle}{\langle n(\Delta \lambda^*_{\rm ob},\Delta z_{\rm ob})\rangle}\,\times\nonumber\\
&\times\int\limits_0^\infty{\rm d}z_{\rm tr}\,\frac{{\rm d}^2 V}{{\rm d} z_{\rm tr}{\rm d}\Omega} \int\limits_0^\infty{\rm d}M\, \hat{w}_{\rm cg}(R^{\rm test},M,z_{\rm tr}) \, \mathcal{S}(M,z_{\rm tr}, \Delta\lambda^*_{\rm ob},\Delta z_{\rm ob})\,,
\end{align}
where $\langle b_{\rm g}\rangle$ is the average galaxy bias, which follows the relation
\begin{equation}\label{eq:b_gal}
\langle b_{\rm g}(\Delta \lambda^*_{\rm ob},\Delta z_{\rm ob})\rangle = b_{{\rm g},0}\,\left(\frac{\bar{\lambda}^*_{\rm ob}}{\lambda^*_{\rm piv}}\right)^{b_{{\rm g},\lambda^*}}\,\left(\frac{1+\bar{z}_{\rm ob}}{1+z_{\rm piv}}\right)^{b_{{\rm g},z}} \,.
\end{equation}
Here, $\lambda^*_{\rm piv}$ and $z_{\rm piv}$ are the pivot values also adopted in Eq.\ \eqref{eq:scalingrelation}, while $\bar{\lambda}^*_{\rm ob}$ and $\bar{z}_{\rm ob}$ are the median values within $\Delta \lambda^*_{\rm ob}$ and $\Delta z_{\rm ob}$, respectively, listed in Table \ref{tab:sample}. By modelling the cluster stacks separately, we verified that Eq.\ \eqref{eq:b_gal} describes the mass proxy and redshift dependence of $b_{\rm g}$  well. In Eq.\ \eqref{eq:w_stack}, $\hat{w}_{\rm cg}$ is the cluster--galaxy correlation function in Eq.\ \eqref{eq:w} where we impose $\langle b_{\rm g}(M,z)\rangle=1$, and $\mathcal{I}$ is the integral constraint, expressed as \citep{Roche1999,Adelberger05,Coupon12}
\begin{align}\label{eq:integral_constraint}
\mathcal{I}(\Delta\lambda^*_{\rm ob},\Delta z_{\rm ob}) =& \,\langle b_{\rm g}(\Delta \lambda^*_{\rm ob},\Delta z_{\rm ob})\rangle \, \times\nonumber\\
&\times\,\frac{\sum_j \mathcal{R}_{\rm c}\mathcal{R}_{\rm g}(R_j|\Delta z_{\rm ob}) \, \langle \tilde{w}_{\rm cg}(R_j,\Delta\lambda^*_{\rm ob},\Delta z_{\rm ob})\rangle}{\sum_j \mathcal{R}_{\rm c}\mathcal{R}_{\rm g}(R_j|\Delta z_{\rm ob})}\,,
\end{align}
where $\mathcal{R}_{\rm c}\mathcal{R}_{\rm g}$ is the normalised number of random cluster--random galaxy pairs, where random redshifts follow the observed distribution within the bin $\Delta z_{\rm ob}$. Moreover, $\langle\tilde{w}_{\rm cg}\rangle$ is obtained by imposing $\mathcal{I}(\Delta\lambda^*_{\rm ob},\Delta z_{\rm ob})=0$ in Eq.\ \eqref{eq:w_stack}, while $j$ runs over the number of cluster-centric radial bins. Here, the minimum radial scale considered is 400 $h^{-1}$kpc, matching the one adopted for the measurements (Sect.\ \ref{sec:measure}). The radial upper limit considered in Eq.\ \eqref{eq:integral_constraint} is derived as
\begin{equation}\label{eq:R_lim}
R_{\rm lim}(z) = D_{\rm l}(z)\,\theta_{\rm lim}\,,
\end{equation}
where $D_{\rm l}$ is the angular diameter distance, computed at the median cluster redshifts listed in Table \ref{tab:sample}, while $\theta_{\rm lim}$ represents the angular extension of the smallest side of the KiDS angular mask. We remark that KiDS is divided into two stripes, with $\theta_{\rm lim}=8.5$ deg on average \citep{Kuijken19}. We adopted this $\theta_{\rm lim}$ value in Eq.\ \eqref{eq:R_lim}, which yields $R_{\rm lim}\simeq70$ $h^{-1}$Mpc for $z_{\rm ob}\in[0.1,0.3)$, $R_{\rm lim}\simeq110$ $h^{-1}$Mpc for $z_{\rm ob}\in[0.3,0.45)$, and $R_{\rm lim}\simeq150$ $h^{-1}$Mpc for $z_{\rm ob}\in[0.45,0.8]$. Assuming the best-fit parameters of the $w_{\rm cg}$ modelling in Eq.\ \eqref{eq:integral_constraint}, we find that $\mathcal{I}(\Delta\lambda^*_{\rm ob},\Delta z_{\rm ob})\simeq10^{-2}$ in all cluster stacks. This is the same order of magnitude as the $w_{\rm cg}$ measurements at large radii and low-to-intermediate redshifts (see Fig.\ \ref{fig:measures_vs_models}).\\
\indent In the reduced shear and correlation function models appearing in Eqs.\ \eqref{eq:g_stack} and \eqref{eq:w_stack}, we assumed the following log-linear model for the $c_{200\rm m}-M_{200\rm m}$ relation:
\begin{equation}\label{eq:cM_rel}
\log c_{200\rm m} = \log c_0 + c_M \log \frac{M}{M_{\rm piv}} + c_z \log \frac{1+z_{\rm tr}}{1+z_{\rm piv}}\,,
\end{equation}
where $M_{\rm piv}$ and $z_{\rm piv}$ are the same as those assumed in Eq.\ \eqref{eq:scalingrelation}. As discussed in Sect.\ \ref{sec:modelling:likelihood}, $\log c_0$ is a free parameter in the analysis, while $c_M$ and $c_z$ are fixed to the fiducial values by \citet{Duffy08}. 

\subsection{Splashback radius estimator}\label{sec:modelling:splashback}
The main goal of this study consists in constraining the following relation \citep[introduced by][]{More15}:
\begin{equation}\label{eq:More15}
\langle \mathcal{R}_{\rm sp}(\Delta\lambda^*_{\rm ob},\Delta z_{\rm ob})\rangle = A_{\rm sp}\,\left[1 + B_{\rm sp}\,{\rm exp}\,\left(-\frac{\langle\nu_{200\rm m}(\Delta\lambda^*_{\rm ob},\Delta z_{\rm ob})\rangle}{2.44}\right)\right]\,,
\end{equation}
where $A_{\rm sp}$ and $B_{\rm sp}$ are free parameters, while $\langle\mathcal{R}_{\rm sp}\rangle$ and $\langle\nu_{200 \rm m}\rangle$ are the average normalised splashback radius and the average peak height, respectively. In our analysis, we derived $\langle\mathcal{R}_{\rm sp}\rangle$ and $\langle\nu_{200 \rm m}\rangle$ posteriors and constrain $A_{\rm sp}$ and $B_{\rm sp}$. Specifically, $\langle\mathcal{R}_{\rm sp}\rangle$ is expressed as
\begin{align}\label{eq:Rsp_estimator}
\langle\mathcal{R}_{\rm sp}(\Delta\lambda^*_{\rm ob},&\Delta z_{\rm ob})\rangle \equiv \left\langle\frac{r_{\rm sp}}{r_{200\rm m}}(\Delta\lambda^*_{\rm ob},\Delta z_{\rm ob})\right\rangle=\nonumber\\
&= \frac{\int\limits_0^\infty{\rm d}z_{\rm tr}\,\frac{{\rm d}^2 V}{{\rm d} z_{\rm tr}{\rm d}\Omega}\int\limits_0^\infty{\rm d}M\, \mathcal{R}_{\rm sp}(M,z_{\rm tr}) \, \mathcal{S}(M,z_{\rm tr}, \Delta\lambda^*_{\rm ob},\Delta z_{\rm ob})}{\langle n(\Delta \lambda^*_{\rm ob},\Delta z_{\rm ob})\rangle}\,,
\end{align}
where
\begin{equation}
\mathcal{R}_{\rm sp}(M,z_{\rm tr}) = \frac{r_{\rm sp}(M,z_{\rm tr})}{r_{200\rm m}(M,z_{\rm tr})}\,.
\end{equation}
Here, we define the splashback radius of a halo as
\begin{equation}\label{eq:Rsp_def}
r_{\rm sp}(M,z_{\rm tr}) \equiv {\rm min}\left[\frac{{\rm d}\log\left[\Delta\rho(r,M,z_{\rm tr})+\rho_{\rm m}(z_{\rm tr})\right]}{{\rm d}\log r}\right]\,,
\end{equation}
where $\Delta\rho(r,M,z_{\rm tr})$ is given by Eq.\ \eqref{eq:DK14}. Equation \eqref{eq:Rsp_def} does not include the contribution by halo miscentring, as it affects scales much smaller than $r_{\rm sp}$ and can be therefore neglected. We evaluated Eq.\ \eqref{eq:Rsp_estimator} at each step of the Markov chain Monte Carlo (MCMC). By replacing $\mathcal{R}_{\rm sp}(M,z_{\rm tr})$ in Eq.\ \eqref{eq:Rsp_estimator} with $\nu_{200\rm m}(M,z_{\rm tr})$, $r_{\rm sp}(M,z_{\rm tr})$, or $M_{200\rm m}$, we derived the expected value of $\nu_{200\rm m}$ (appearing in Eq.\ \ref{eq:More15}), $r_{\rm sp}$, or $M_{200\rm m}$, respectively, in bins of $\lambda^*_{\rm ob}$ and $z_{\rm ob}$. \\
\indent We emphasise that Eq.\ \eqref{eq:Rsp_estimator} removes any dependence on the radial binning scheme. While stacking clusters as a function of $r/r_{\rm 200m}$ could in principle enhance the signal-to-noise of $r_{\rm sp}$ \citep{Diemer14}, this approach is compromised for observational data due to noisy $r_{\rm 200m}$ estimates. Equation \eqref{eq:Rsp_estimator} solves this issue, as it models the average profile of a reconstructed population of dark matter haloes. Conversely, in simulations, average profiles of cluster ensembles are modelled as a single profile, thanks to the reduced scatter on $r_{200\rm m}$.

\subsection{Likelihood and priors}\label{sec:modelling:likelihood}
\begin{figure*}[t!]
\centering\includegraphics[width = \hsize-1.5cm, height = 11.5cm] {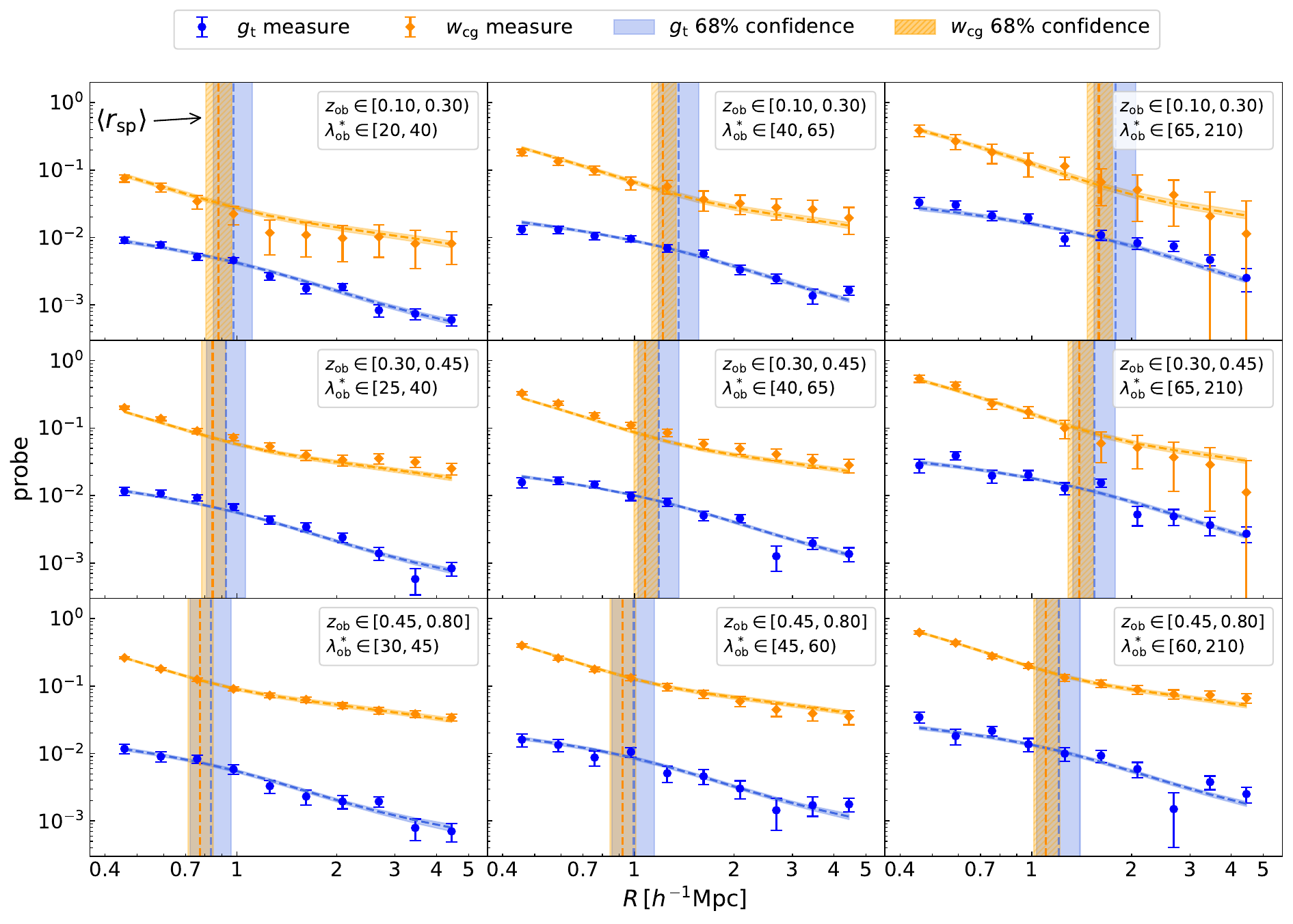}
\caption{Measurements of $g_{\rm t}$ (blue dots) and $w_{\rm cg}$ (orange diamonds) profiles of the AMICO KiDS-1000 galaxy clusters, in bins of $z$ (increasing from top to bottom) and $\lambda^*$ (increasing from left to right). The error bars are the sum of statistical errors and residual uncertainties from systematic errors (see Sect.\ \ref{sec:modelling:likelihood}). The bands superimposed onto the measurements represent the 68\% confidence levels of the $g_{\rm t}$ (blue) and $w_{\rm cg}$ (orange) models. The vertical bands show the 68\% confidence of the splashback radius, derived from the modelling of $g_{\rm t}$ (blue) and $w_{\rm cg}$ (orange). The dashed lines represent median values.}
\label{fig:measures_vs_models}
\end{figure*}
In this work, we analysed $g_{\rm t}$ and $w_{\rm cg}$ separately to compare two independent estimates of $r_{\rm sp}$. For both these probes, we performed a joint Bayesian analysis of all the cluster stacks, by means of an MCMC algorithm. The likelihood function of a probe is expressed as
\begin{equation}
\mathcal{L} = \prod_{i=1}^{N_{\lambda^*}}\prod_{j=1}^{N_{z}} \mathcal{L}_{ij}\,,
\end{equation}
where $N_{\lambda^*}$ and $N_{z}$ are the numbers of cluster intrinsic richness and redshift bins used for the stacked measurements, respectively, while $\mathcal{L}_{ij}$ is a Gaussian likelihood function defined as
\begin{equation}
\mathcal{L}_{ij} \propto \exp(-\chi_{ij}^2/2)\,,
\end{equation}
with
\begin{align}
&\chi_{ij}^2=\sum_{k=1}^{N_R}\sum_{l=1}^{N_R}
  \left(\mathcal{O}_{ijk}-\mathcal{M}_{ijk} \right)\, C_{ijkl}^{-1}\, \left(
  \mathcal{O}_{ijl}-\mathcal{M}_{ijl}\right)\,.
\end{align}
Here, $\mathcal{O}_{ijk}$ is either the stacked weak lensing (Eq.\ \ref{eq:DeltaSigma_measure_stack}) or the cluster--galaxy correlation (Eq.\ \ref{eq:LS_estimator}) measurement, while $\mathcal{M}_{ijk}$ is the corresponding model, namely Eq.\ \eqref{eq:g_stack} or Eq.\ \eqref{eq:w_stack}, respectively. The indices $k$ and $l$ run over the number of radial bins, $N_R$, and $C_{ijkl}^{-1}$ is the inverse of the covariance matrix. Following \citet{Hartlap07}, we applied the standard correction when inverting the covariance matrix. In particular, $C_{ijkl}$ is defined as
\begin{equation}
C_{ijkl} = C_{ijkl}^{\rm stat} + C_{ijkl}^{\rm sys}\,,
\end{equation}
where $C_{ijkl}^{\rm stat}$ is the statical part of the covariance, estimated through resampling as discussed in Sect.\ \ref{sec:measure}, while $C_{ijkl}^{\rm sys}$ accounts for residual uncertainties on systematic errors that are not included in the model. In the case of $g_{\rm t}$, this covariance contribution is written as
\begin{equation}\label{eq:C_sys_g}
C_{ijkl}^{{\rm sys},g_{\rm t}} = (\sigma_m^2 + \sigma_{\rm\scriptscriptstyle SOM}^2 + \sigma_{\rm\scriptscriptstyle OP}^2)\, g_{\rm t,k}^{\rm ob}(\Delta\lambda^*_{{\rm ob},i},\Delta z_{{\rm ob},j})\,g_{\rm t,l}^{\rm ob}(\Delta\lambda^*_{{\rm ob},i},\Delta z_{{\rm ob},j})\,,
\end{equation}
where $g_{\rm t}^{\rm ob}$ is the observed stacked reduced shear, $\sigma_m=0.02$ is the uncertainty on the multiplicative shear bias, corresponding to the largest multiplicative bias uncertainty reported across the tomographic bins analysed by \citet{Giblin21}. Furthermore, $\sigma_{\rm\scriptscriptstyle SOM}$ accounts for the uncertainty on the SOM-reconstructed background redshift distributions, amounting to $\sigma_{\rm\scriptscriptstyle SOM}=0.01$ for the first two cluster redshift bins and to $\sigma_{\rm\scriptscriptstyle SOM}=0.04$ for the last one \citepalias[see Sect.\ 4.4 in][]{Lesci25}, while $\sigma_{\rm\scriptscriptstyle OP}=0.03$ is the residual uncertainty due to orientation and projection effects \citepalias[see Sect.\ 5.4 in][]{Lesci25}. \\
\indent Measurements of $w_{\rm cg}$ are not affected by multiplicative shear biases. We also ignored orientation and projection effects in this case, assuming that their impact is absorbed by the free galaxy bias in Eq.\ \eqref{eq:w_stack}. The impact of this approach on galaxy bias constraints shall be tested in future studies. Conversely, we included the contribution to $w_{\rm cg}$ measurements due to the uncertainty on the mean of the SOM-reconstructed $n(z_{\rm g})$, amounting to 2\% \citep{Hildebrandt21}. To this end, we estimated $\langle w_{\rm cg}(R,\Delta\lambda^*_{\rm ob},\Delta z_{\rm ob})\rangle$ in Eq.\ \eqref{eq:w_stack} at the model best-fit parameters, using the reconstructed $n(z_{\rm g})$ from Sect.\ \ref{sec:dataset} in Eq.\ \eqref{eq:w}. We then repeated this process while shifting $n_{\rm bkg}(z_{\rm g})$ by $\pm2\%$ of its mean. This yielded a relative uncertainty on $w_{\rm cg}$ that is approximately constant with $R$ and $\lambda^*_{\rm ob}$, amounting to 3\% for the first two cluster redshift bins and 2\% for the last one. Thus, the residual systematics term in the case of $w_{\rm cg}$ has the following expression:
\begin{equation}
C_{ijkl}^{{\rm sys},w_{\rm cg}} = \sigma_{\rm\scriptscriptstyle SOM}^2\, \omega_{{\rm cg},k}^{\rm ob}(\Delta\lambda^*_{{\rm ob},i},\Delta z_{{\rm ob},j})\,\omega_{{\rm cg},l}^{\rm ob}(\Delta\lambda^*_{{\rm ob},i},\Delta z_{{\rm ob},j})\,,
\end{equation}
where $w_{\rm cg}^{\rm ob}$ is the observed cluster--galaxy correlation function, $\sigma_{\scriptscriptstyle\rm SOM}=0.03$ for $z_{\rm ob}\in[0.1,0.3)$ and $z_{\rm ob}\in[0.3,0.45)$, while $\sigma_{\scriptscriptstyle\rm SOM}=0.02$ for $z_{\rm ob}\in[0.45,0.8]$. \\
\indent The parameter priors adopted in the analysis are reported in Table \ref{tab:priors_and_posteriors}. We assumed a Gaussian prior on the truncation factor of the one-halo term, $F_{\rm t}$, in Eq.\ \eqref{eq:DK14}, with mean 1.495 \citepalias[following][]{Diemer14} and standard deviation 0.3 \citepalias[corresponding to that assumed by][in the case of $M_{200\rm m}$]{Lesci25}, also imposing  $F_{\rm t}>0$. Gaussian priors were also adopted on the parameters of the transition factor (Eq.\ \ref{eq:DK14:transition}), namely $\gamma_0$, appearing in the relation $\gamma=\gamma_0\nu_{\rm vir}$ (see Sect.\ \ref{sec:modelling:single}), and $\beta$; both have a mean of 4 \citepalias[following][]{Diemer14} and a standard deviation of 1.6 \citep[following][]{More16,Shin19,Murata20,Rana23}. For the outer profile parameters (Eq.\ \ref{eq:DK14:infall}), as introduced in Sect.\ \ref{sec:modelling:single}, we assumed that $s_{\rm e}$ is independent of mass and redshift, and adopted a flat prior on this parameter. On the other hand, $b_{\rm e}$ varies with redshift (Eq.\ \ref{eq:b_e}), and flat priors are adopted for its amplitude, $b_{{\rm e},0}$, and redshift scaling, $b_{{\rm e},z}$.\\
\indent The fraction of miscentred clusters, $f_{\rm off}$, in Eq.\ \eqref{eq:g_tot} is assigned a Gaussian prior with mean 0.3 and standard deviation 0.1, also imposing  $f_{\rm off} > 0$, while the miscentring scale, $\sigma_{\rm off}$, in Eq.\ \eqref{eq:Poff} is given the uniform prior $[0, 0.5]$ $h^{-1}$Mpc. These priors are motivated by previous studies of centre offsets in simulations \citep{Yan20,Sommer23} and observations across various mass density tracers \citep{Saro15,Zhang19,Seppi23,Ding24}. In the $g_{\rm t}$ modelling, we assumed flat priors on the $\log\lambda^*-\log M_{200\rm m}$ scaling relation parameters (Eqs.\ \ref{eq:scalingrelation_PDF} and \ref{eq:scalingrelation}), namely $A$, $B$, $C$, and $\sigma_{\rm intr}$. On the other hand, to analyse $w_{\rm cg}$ we assumed the $g_{\rm t}$ posteriors on these parameters as priors. A flat prior was used for the amplitude of the $\log c_{200\rm m}-\log M_{200\rm m}$ relation (Eq.\ \ref{eq:cM_rel}). In this relation, we fixed $c_M=-0.107$ and $c_z=-1.16$ \citep[following][]{Duffy08}. We expect that our priors on $\sigma_{\rm intr}$ and $c_{200\rm m}$ account for theoretical uncertainties in baryonic effects on cluster profiles \citep{Schaller15,Henson17,Lee18,Beltz-Mohrmann21,Shao24}. \\
\indent In the $w_{\rm cg}$ modelling, we assumed flat priors on $b_{{\rm g},0}$, $b_{{\rm g},\lambda^*}$, and $b_{{\rm g},z}$, which enter the relation between the average galaxy bias, $\langle b_{\rm g} (\Delta\lambda^*_{\rm ob},\Delta z_{\rm ob})\rangle$, and cluster observables (Eq.\ \ref{eq:b_gal}). The $\mathcal{B}_{\rm \scriptscriptstyle HMF}$ factor in Eq.\ \eqref{eq:B_HMF} provides a correction to the standard \citet{Tinker08} halo mass function. For the parameters $s$ and $q$, we employed the \citet{Costanzi19} prescription, assigning a bivariate Gaussian prior with mean values $(0.037, 1.008)$ and covariance matrix $C_{\rm \scriptscriptstyle HMF}$ expressed as
\begin{equation}\label{covMtinker}
C_{\rm \scriptscriptstyle HMF} = 
\begin{pmatrix}
0.00019 & 0.00024 \\
0.00024 & 0.00038
\end{pmatrix}
\,.
\end{equation}
Although baryonic physics uncertainties are not taken into account in this prescription, they are negligible compared to the precision of the $\mathcal{B}_{\rm \scriptscriptstyle HMF}$ measurement \citep[see][and references therein]{Costanzi19}.

\section{Results}\label{sec:results}
\begin{figure*}[t!]
\centering\includegraphics[width = \hsize, height = 13.5cm] {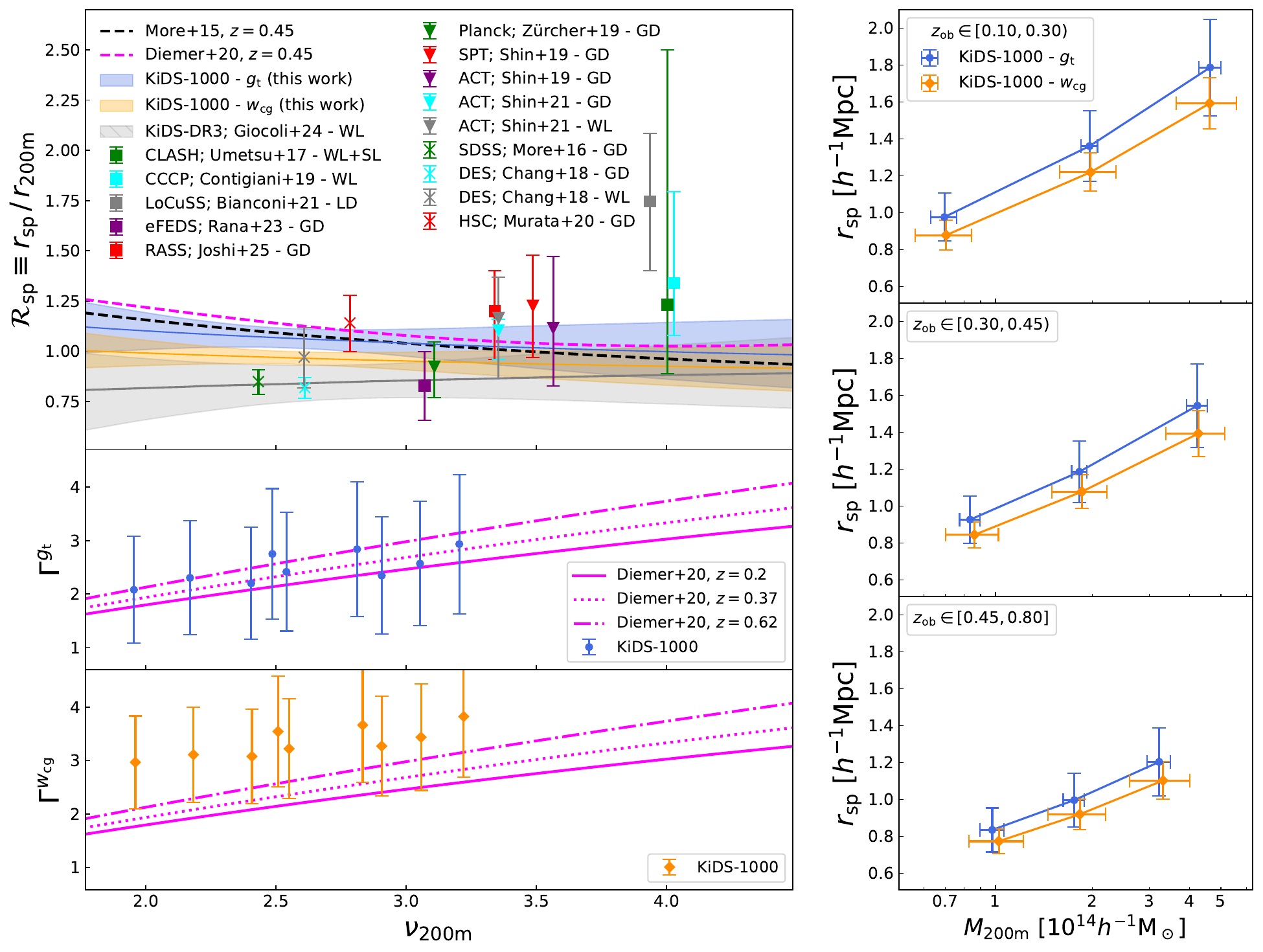}
\caption{\textit{Top left}: Constraints on the ratio of $r_{\rm sp}$ to $r_{200\rm m}$ as a function of $\nu_{200\rm m}$  obtained from the modelling of $g_{\rm t}$ (blue band) and $w_{\rm cg}$ (orange band) presented in this work, and by \citet{Giocoli24} in KiDS-DR3 (grey band). The median theoretical models from \citet[dashed black line]{More15} and \citet[dashed magenta line]{Diemer20_2} are shown. Both models are computed at $z=0.45$. Squares represent the results from X-ray-selected clusters by \citet[][green]{Umetsu17}, \citet[][cyan]{Contigiani19}, \citet[][grey]{Bianconi21}, \citet[][purple]{Rana23}, and \citet[][red]{Joshi25}. Triangles display results from SZ-selected cluster catalogues from \citet[][green]{Zurcher19}, \citet[][red and purple]{Shin19}, and \citet[][cyan and grey]{Shin21}. Crosses show the results from optically selected clusters from \citet{More16} and \citet[][green]{Baxter17}, \citet[][cyan and grey]{Chang18}, and \citet[][red]{Murata20}. The probes used in these analyses, namely weak lensing (WL), strong lensing (SL), luminosity distribution (LD), and galaxy distribution (GD), are reported in the legend. \textit{Centre left}: Mass accretion rates from $g_{\rm t}$ measurements. \textit{Bottom left}: Mass accretion rates from $w_{\rm cg}$ measurements. Here, the model from \citet{Diemer20_2} is shown, computed at $z=0.2$ (solid lines), $z=0.37$ (dotted lines), and $z=0.62$ (dash-dotted lines). \textit{Right panels}: $r_{\rm sp}$ derived from the modelling of $g_{\rm t}$ (blue dots) and $w_{\rm cg}$ (orange diamonds), as a function of $M_{200\rm m}$ and for $z_{\rm ob}\in[0.1,0.3)$ (\textit{top}), $z_{\rm ob}\in[0.3,0.45)$ (\textit{middle}), and $z_{\rm ob}\in[0.45,0.8]$ (\textit{bottom}).}
\label{fig:rsp_r200m}
\end{figure*}
\begin{figure}[t]
\centering\includegraphics[width = \hsize-0.5cm, height = 6.cm] {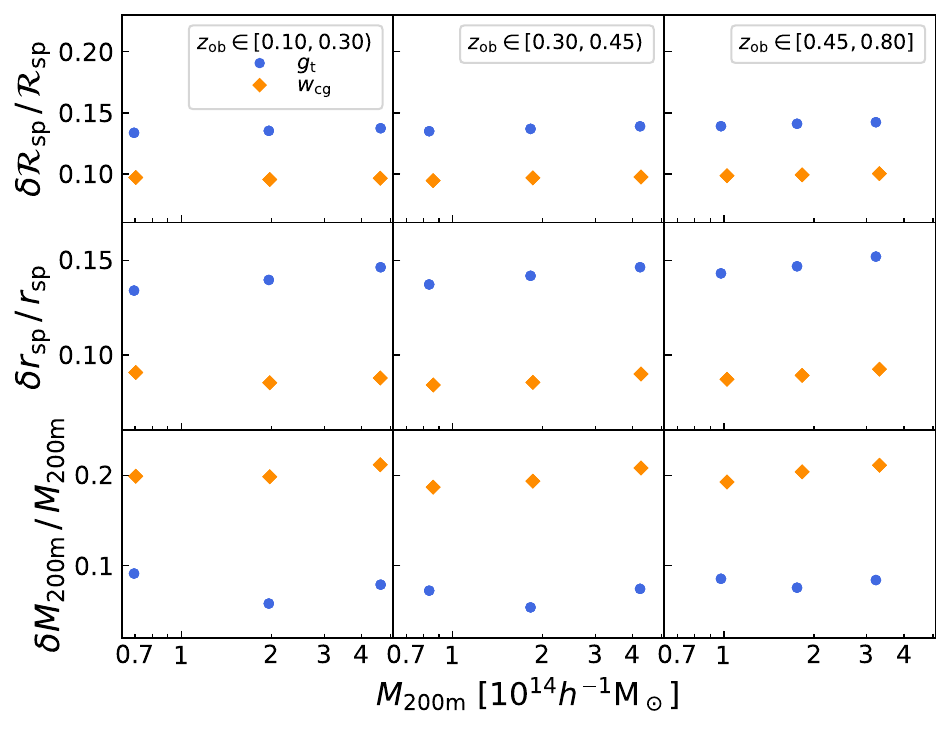}
\caption{Precision of the $\mathcal{R}_{\rm sp}$ (\textit{top row}), $r_{\rm sp}$ (\textit{middle row}), and $M_{200\rm m}$ (\textit{bottom row}) constraints, obtained from the modelling of $g_{\rm t}$ (blue dots) and $w_{\rm cg}$ (orange diamonds), for $z_{\rm ob}\in[0.1,0.3)$ (\textit{left column}), $z_{\rm ob}\in[0.3,0.45)$ (\textit{central column}), and $z_{\rm ob}\in[0.45,0.8]$ (\textit{right column}).}
\label{fig:precision}
\end{figure}
In the modelling presented here, we adopted a flat $\Lambda$CDM cosmological model assuming the median values of the posteriors derived by \citetalias{Lesci25}, namely $\Omega_{\rm m}=0.22$ and $\sigma_8=0.86$. The other cosmological parameters were fixed to the results by \citet[][Table 2, TT, TE + EE + lowE + lensing, referred to as Planck18 hereafter]{Planck18}. Figure \ref{fig:measures_vs_models} shows the $g_{\rm t}$ and $w_{\rm cg}$ measurements against the corresponding best-fit models. We obtained a reduced $\chi^2$ of $\chi_{\rm r}^2 \simeq 1.2$ for $g_{\rm t}$ and $\chi_{\rm r}^2 \simeq 1.0$ for $w_{\rm cg}$. The $\langle r_{\rm sp}\rangle$ estimates are shown as vertical bands, and are derived by computing the model described in Sect.\ \ref{sec:modelling:splashback} at each MCMC step. The results presented in this section on splashback radii, masses, and other derived quantities are also reported in Table \ref{tab:sample}. The robustness of the results against our modelling choices is discussed in Appendix \ref{app:robustness}. \\
\indent The top-left panel of Fig.\ \ref{fig:rsp_r200m} displays our constraints on $\langle\mathcal{R}_{\rm sp}\rangle\equiv\langle r_{\rm sp}/r_{200\rm m}\rangle$, based on the methods outlined in Sect.\ \ref{sec:modelling:splashback}. Specifically, we first obtained the posteriors on $\langle\mathcal{R}_{\rm sp}\rangle$ and $\langle\nu_{200\rm m}\rangle$ in each stack using Eq.\ \eqref{eq:Rsp_estimator}, and then constrained $A_{\rm sp}$ and $B_{\rm sp}$ in Eq.\ \eqref{eq:More15} through an MCMC algorithm. In this process, we ignored the statistical uncertainty on $\langle\nu_{200\rm m}\rangle$, corresponding to about 2\%. We obtained the following constraints:
\begin{align}\label{eq:results:Rsp}
&A_{\rm sp}^{g_{\rm t}}=0.91^{+0.30}_{-0.30},\,\,\,\,\,B_{\rm sp}^{g_{\rm t}}=0.49^{+1.63}_{-0.84}\nonumber\\
&A_{\rm sp}^{w_{\rm cg}}=0.87^{+0.20}_{-0.21},\,\,\,B_{\rm sp}^{w_{\rm cg}}=0.30^{+0.97}_{-0.59}\,,
\end{align}
where the $g_{\rm t}$ and $w_{\rm cg}$ superscripts represent the results derived from the modelling of weak-lensing and cluster--galaxy correlation function measurements, respectively. As we can see in the top-left panel of Fig.\ \ref{fig:rsp_r200m}, our estimates of $\mathcal{R}_{\rm sp}$ derived from these two probes are consistent within $1\sigma$. This agreement between the two probes is in line with the findings of previous studies \citep{Chang18,Shin21}. Furthermore, as reported in Table \ref{tab:sample}, we find no significant trends of $\mathcal{R}_{\rm sp}$ with redshift or intrinsic richness, indicating that $r_{\rm sp}$ and $r_{200\rm m}$ share similar dependences on $M_{200\rm m}$ and $z$, in agreement with simulation predictions \citep{More15,Diemer20_2}. \\
\indent Compared to the $\mathcal{R}_{\rm sp}$ constraints from \citet{Giocoli24} shown in Fig.\ \ref{fig:rsp_r200m}, based on weak-lensing measurements in KiDS-DR3, our $g_{\rm t}$ modelling yields results that agree within $1\sigma$ for $\nu_{200\rm m}>3$, reducing to $2\sigma$ agreement at lower $\nu_{200\rm m}$. We verified the consistency within $1\sigma$ between our $r_{\rm sp}$ values and those from \citet{Giocoli24}. Furthermore, our mass constraints agree with those by \citetalias{Lesci25} (see Appendix \ref{app:robustness}). Thus, we conclude that the observed small shift in $\mathcal{R}_{\rm sp}$ aligns with the cluster mass discrepancy between KiDS-1000 and KiDS-DR3 reported by \citetalias{Lesci25}. In that work, the KiDS-1000 analysis yields lower masses at the low-mass end. These lower masses imply smaller $r_{200\rm m}$ values, which in turn explain the larger $\mathcal{R}_{\rm sp}$ we find compared to \citet{Giocoli24}. \citetalias{Lesci25} ascribe these mass differences to improved systematic uncertainty modelling in KiDS-1000 (see their Appendix C). The modelling of $w_{\rm cg}$ provides $\mathcal{R}_{\rm sp}$ constraints in better agreement with \citet{Giocoli24}. As discussed below, this agreement originates mainly from the lower $r_{\rm sp}$ values derived from cluster--galaxy correlation function measurements, which we attribute to dynamical friction. Indeed, $g_{\rm t}$ probes the total matter distribution and is insensitive to galaxy-specific dynamical processes, while $w_{\rm cg}$ traces the spatial distribution of satellite galaxies. Dynamical friction acts on these satellites, causing them to lose kinetic energy and sink inwards. This shifts the splashback radius traced by $w_{\rm cg}$ to smaller values. \\
\indent The precision of the $\mathcal{R}_{\rm sp}-\nu_{200\rm m}$ relation parameters, reported in Eq.\ \eqref{eq:results:Rsp}, aligns with the findings of \citet{Giocoli24}. Our analysis, however, incorporates a more comprehensive marginalisation over statistical uncertainties, including those from the transition factor parameters (Eq.\ \ref{eq:DK14:transition}), the halo miscentring (Sect.\ \ref{sec:modelling:single_probes:g}), the intrinsic scatter of the $\log\lambda^*-\log M_{200\rm m}$ relation (Eq.\ \ref{eq:scalingrelation_PDF}), and the halo mass function correction (Eqs.\ \ref{eq:B_HMF} and \ref{covMtinker}). We also applied more conservative $\lambda^*$ cuts. \\ 
\indent The $\mathcal{R}_{\rm sp}$ models in the top-left panel of Fig.\ \ref{fig:rsp_r200m} are computed across the mass range of our sample, $M_{200\rm m} \in [0.4,20]$ $10^{14} h^{-1} M_\odot$, which is derived from individual cluster mass estimates based on Eq. (54) in \citetalias{Lesci25} and on the parameter posteriors obtained from the $g_{\rm t}$ modelling (Table \ref{tab:priors_and_posteriors}). The theoretical models shown in the figure, by \citet{More15} and \citet{Diemer20_2}, are computed at $z=0.45$, which corresponds to the midpoint of the redshift range covered by our cluster sample. We remark that the redshift dependence of these models is very mild. Our constraints are in good agreement with these theoretical predictions. The largest deviations appear for the $w_{\rm cg}$ results at $\nu_{\rm 200m}<3$, reaching 2$\sigma$ with \citet{Diemer20_2} at $\nu_{\rm 200m}\sim2$. \\
\indent The top-left panel of Fig.\ \ref{fig:rsp_r200m} also shows the comparison with literature results based on X-ray \citep{Umetsu17,Contigiani19,Bianconi21,Rana23,Joshi25}, Sunyaev-Zeldovich \citep[SZ;][]{Zurcher19,Shin19,Shin21}, and optical \citep{More16,Baxter17,Chang18,Murata20} samples of galaxy clusters.\footnote{The legend in the top-left panel of Fig.\ \ref{fig:rsp_r200m} reports the surveys these works are based on, namely the Cluster Lensing And Supernova survey with Hubble \citep[CLASH;][]{Postman12}, the Canadian Cluster Comparison Project \citep[CCCP;][]{Hoekstra12}, the Local Cluster Substructure Survey \citep[LoCuSS;][]{Bohringer04}, the eROSITA final equatorial depth survey \citep[eFEDS;][]{Bulbul22}, the ROSAT All-Sky Survey \citep[RASS;][]{Voges1999}, Planck \citep{Planck18}, the South Pole Telescope \citep[SPT;][]{Bleem15} survey, the Atacama Cosmology Telescope \citep[ACT;][]{Hilton18} survey, the Sloan Digital Sky Survey \citep[SDSS;][]{Aihara11}, the Dark Energy Survey \citep[DES;][]{DES05}, and the Hyper Suprime-Cam \citep[HSC;][]{Miyazaki12} survey.} We find general agreement with these literature results. The largest discrepancies are with the results from \citet{More16} and \citet{Chang18}, which are based on cluster member galaxy distributions. This may in part be explained  by dynamical friction, which has a greater effect on the galaxy samples used in those studies. In fact, \citet{More16} and \citet{Chang18} used shallower galaxy samples, with a magnitude limit of $i<21.5$. Compared to our work, this implies a selection of more massive galaxies, for which the impact of dynamical friction is stronger. Future photometric surveys will allow for a detailed investigation of the effects by dynamical friction. \\
\indent As displayed in the top and middle panels of Fig.\ \ref{fig:precision}, the precision on $\mathcal{R}_{\rm sp}$ and $r_{\rm sp}$ derived from the modelling of $g_{\rm t}$ and $w_{\rm cg}$ is constant with $z$ and $M_{200\rm m}$. It amounts to 14\% for $g_{\rm t}$ and to 10\% for $w_{\rm cg}$. These precisions are among the best available in the literature to date. Furthermore, these results show that $w_{\rm cg}$ is a stronger probe of the splashback radius, as it directly constrains the infalling term parameters, which are not accessible via weak-lensing observations (see Appendix \ref{app:nuisance}). \\
\indent The right panels of Fig.\ \ref{fig:rsp_r200m} show that $r_{\rm sp}$ estimates from $w_{\rm cg}$ are systematically lower than those from the $g_{\rm t}$ modelling, although the values remain consistent within $1\sigma$. This small offset can be attributed to dynamical friction affecting cluster members \citep{Diemer17,Chang18}, which also causes haloes to appear more concentrated. Figure \ref{fig:rsp_r200m} also displays an agreement on the $M_{200\rm m}$ constraints derived from the two probes. As we detail in Appendix \ref{app:nuisance}, the larger $M_{200\rm m}$ uncertainties derived from $w_{\rm cg}$, displayed in the bottom panels of Fig.\ \ref{fig:precision}, are attributed to the differences in the constraints on the $A$ and $\sigma_{\rm intr}$ parameters, appearing in the $\log\lambda^*-\log M_{200\rm m}$ relation. The right panels of Fig.\ \ref{fig:rsp_r200m} also show that $r_{\rm sp}$ mildly decreases with redshift, as higher redshifts correspond to larger halo mass accretion rates, leading to smaller splashback radii \citep{Diemer14,Diemer17}. \\
\indent \citet{More15} derived the following expression for the dimensionless mass accretion rate, $\Gamma$:
\begin{equation}\label{eq:Gamma}
\Gamma \equiv \frac{\Delta\log{M_{\rm vir}}}{\Delta\log a} = 0.935 - 3.04 \ln\left(\frac{\mathcal{R}_{\rm sp}}{0.54 + 0.286\,\Omega_{\rm m}(z)} - 1\right) \,,
\end{equation}
where $a$ is the scale factor. We computed $\Gamma$ posteriors by injecting $\langle\mathcal{R}_{\rm sp}(\Delta\lambda^*_{\rm ob},\Delta z_{\rm ob})\rangle$ values, derived from Eq.\ \eqref{eq:Rsp_estimator} at each MCMC step, into Eq.\ \eqref{eq:Gamma}. Furthermore, $\Omega_{\rm m}(z)$ in Eq.\ \eqref{eq:Gamma} is computed at the median cluster redshifts reported in Table \ref{tab:sample}. As shown in the middle and bottom left panels of Fig.\ \ref{fig:rsp_r200m}, our constraints on $\Gamma^{g_{\rm t}}$ and $\Gamma^{w_{\rm cg}}$, derived from the modelling of $g_{\rm t}$ and $w_{\rm cg}$, respectively, agree with the theoretical model by \citet{Diemer20_2}. This agreement is expected, as our $\mathcal{R}_{\rm sp}$ results are consistent with the $\Lambda$CDM predictions from \citet{More15} and \citet{Diemer20_2}, and Eq.\ \eqref{eq:Gamma} is itself derived from those same simulations.\\
\indent As reported in Table \ref{tab:priors_and_posteriors} and detailed in Appendix \ref{app:nuisance}, the $w_{\rm cg}$ modelling yields robust constraints on the \citetalias{Diemer14} infalling term parameters, namely $b_{\rm e}$ and $s_{\rm e}$, while weak-lensing observations do not probe this region. This is not attributable to different measurement precisions, which are similar for the two probes, but to the fact that $w_{\rm cg}$ is more directly sensitive to the spatial matter distribution in the infall region. Weak lensing, instead, probes $\Delta\Sigma_{\rm t}$, an enclosed-mass quantity that is less sensitive to the detailed structure of the infalling region. Further discussions on the constraints on \citetalias{Diemer14} parameters can be found in Appendix \ref{app:nuisance}. In agreement with \citetalias{Lesci25}, we find that the $f_{\rm off}$ posterior aligns with the prior and $\sigma_{\rm off}=0.21\pm0.11$ $h^{-1}$Mpc. The latter constraint also agrees  with literature simulation and observational results \citep{Saro15,Zhang19,Yan20,Seppi23,Sommer23}. As shown in Table \ref{tab:priors_and_posteriors}, we constrain the average galaxy bias parameters in Eq.\ \eqref{eq:b_gal} through the $w_{\rm cg}$ modelling, obtaining $b_{{\rm g},0}=1.17^{+0.18}_{-0.17}$, $b_{{\rm g},\lambda^*}=-0.10^{+0.11}_{-0.11}$, and $b_{{\rm g},z}=0.31^{+0.35}_{-0.33}$. A positive redshift evolution of $\langle b_{\rm g}\rangle$ is expected, because galaxies at higher redshifts reside in rarer, more biased haloes. Our analysis, however, does not yield a significant redshift trend, as $b_{{\rm g},z}$ is consistent with zero within $1\sigma$. Similarly, we find no significant dependence on $\lambda^*$. Future, more precise $w_{\rm cg}$ measurements will allow for a more detailed modelling of the galaxy bias.

\section{Discussion and conclusions}\label{sec:conclusions}
In this paper we have presented an analysis of the splashback radius for the richest 8730 optically selected galaxy clusters from the AMICO KiDS-1000 catalogue \citep{Maturi25}. The sample covers an effective area of 839 deg$^2$ and spans the redshift range $z\in[0.1,0.8]$. We measured and modelled the reduced tangential shear, $g_{\rm t}$, and the cluster--galaxy projected correlation function, $w_{\rm cg}$, of cluster ensembles, binning by redshift and mass proxy. Our modelling strategy, based on the techniques employed in recent cosmological analyses, allowed us to reconstruct the average properties of the underlying dark matter halo population, such as the normalised splashback radius, $\mathcal{R}_{\rm sp}$, the splashback radius, $r_{\rm sp}$, and the mass accretion rate, $\Gamma$. \\
\indent Our constraints confirm the results of $\Lambda$CDM simulations and agree with observational studies in the literature. The precision of our $r_{\rm sp}$ estimates is among the best available in the literature to date. It amounts to 14\% in the case of $g_{\rm t}$ modelling, while $w_{\rm cg}$ constraints have a precision of 10\%. The measurements of $w_{\rm cg}$ also uniquely constrain the profile of the infalling material, which is not probed by weak-lensing observations.\\
\indent The $r_{\rm sp}$ estimates from $g_{\rm t}$ and $w_{\rm cg}$ agree within $1\sigma$, although $w_{\rm cg}$ tends to yield slightly lower values. This offset is consistent with expectations from dynamical friction, which preferentially affects massive satellite galaxies and can bias $w_{\rm cg}$ measurements, especially in the presence of an incomplete galaxy sample. In contrast, $g_{\rm t}$ traces the total matter distribution and is therefore not subject to this effect. Given this physical distinction, the two probes cannot be considered equivalent tracers of $r_{\rm sp}$: $w_{\rm cg}$ may provide a biased estimate, whereas $g_{\rm t}$ is expected to be unbiased. For this reason, we did not combine the two measurements into a single constraint, as this would effectively average biased and unbiased estimates. Future analyses, based on more complete galaxy samples or incorporating an explicit modelling of dynamical friction, may enable a consistent joint determination of $r_{\rm sp}$. We also note that recent studies based on Stage-III surveys \citep[e.g.][]{Zurcher19,Murata20} were unable to constrain any trend in the impact of dynamical friction on the splashback radius as a function of limiting magnitude, due to insufficiently large datasets. \\
\indent As discussed in Sect.\ \ref{sec:modelling:likelihood}, we assumed that the net effect of orientation and projections on the $w_{\rm cg}$ measurements is absorbed by the free galaxy bias. The impact of this assumption on $w_{\rm cg}$ will be tested in future studies based on simulations. Intrinsic alignments are expected to have a negligible impact in our analysis, contributing only a sub-percent systematic error to cluster profiles \citep{Chisari14,Sereno18}. In the modelling of $g_{\rm t}$, \citetalias{Lesci25} demonstrate that anisotropic boosts, which affect the correlation functions on large scales due to optical projection effects \citep{Sunayama20,Wu22,Sunayama23,Park23,Zhou24,Nde25}, have a negligible impact on our mass constraints. In particular, \citetalias{Lesci25} adopted a model designed for cluster samples selected with the red-sequence matched-filter probabilistic percolation (redMaPPer) algorithm \citep[][]{Rykoff14}. This approximation holds in the case of a cosmological two-halo term, which mildly depends on the free parameters considered in the analysis. In this work, the cosmological two-halo was replaced by a free infalling term (Eq.\ \ref{eq:DK14:infall}), whose shape is poorly constrained by weak-lensing observations. Thus, we do not expect the anisotropic boosts to have a dramatic impact on the results from the $g_{\rm t}$ modelling. Nonetheless, the constraints obtained from $w_{\rm cg}$ measurements may be biased due to this selection effect. In the future we will effectively test the impact of this selection bias on $r_{\rm sp}$ by developing dedicated mock datasets that simulate AMICO detections. We will also test how the assumption of the \citet{Diemer25} model, adopted as an alternative to the \citetalias{Diemer14} profile (Eq.\ \ref{eq:DK14}), affects our results. The \citet{Diemer25} model requires higher data quality because it includes a larger number of free parameters compared to \citetalias{Diemer14}, and KiDS-Legacy data \citep{Wright24} can provide the statistical power needed to constrain these additional degrees of freedom. We also note that the \citetalias{Diemer14} model is extrapolated up to very large scales for the derivation of $g_{\rm t}$ (Eq.\ \ref{eq:g_tot}) and $w_{\rm cg}$ (Eq.\ \ref{eq:w}) models. The impact of such an extrapolation will be assessed through simulations. \\
\indent Based on simulations, \citet{Shin23} demonstrate that the luminosity difference between the brightest cluster galaxy and its brightest satellites, known as the magnitude gap, serves as a reliable tracer of both the mass accretion rate and $r_{\rm sp}$. This gap widens as satellites undergo disruption and halo accretion slows, and therefore also serves as an indicator of system age \citep[][]{Jones03,Deason13,Yang25}. Data from KiDS-Legacy and Stage-IV surveys, such as \textit{Euclid} \citep{Mellier25} and the \textit{Vera C. Rubin Observatory} \citep[\textit{Rubin}/LSST;][]{LSST}, will allow us to empirically constrain the correlation between $r_{\rm sp}$ and the magnitude gap. Stage-IV data can potentially also shed light on the stellar splashback radius, traced by the intracluster light \citep{Dacunha25,Walker25}, and on eventual deviations from general relativity in the outskirts of galaxy clusters \citep{Contigiani23,Butt25}. As discussed in Appendix \ref{app:robustness}, $r_{\rm sp}$ estimates obtained assuming the \citetalias{Planck18} cosmology are systematically lower than those from the baseline analysis, though they remain consistent within 0.5$\sigma$. With future data, the reduction in statistical uncertainties will increase the $r_{\rm sp}$ sensitivity to cosmological parameters. Lastly, while the present $r_{\rm sp}$ measurements from $g_{\rm t}$ and $w_{\rm cg}$ are consistent within $1\sigma$, upcoming data may better constrain the impact of dynamical friction on cluster galaxies.

\begin{acknowledgements}
Based on observations made with ESO Telescopes at the La Silla Paranal Observatory under programme IDs 177.A-3016, 177.A-3017, 177.A-3018 and 179.A-2004, and on data products produced by the KiDS consortium. The KiDS production team acknowledges support from: Deutsche Forschungsgemeinschaft, ERC, NOVA and NWO-M grants; Target; the University of Padova, and the University Federico II (Naples). We acknowledge the financial contribution from the grant PRIN-MUR 2022 20227RNLY3 “The concordance cosmological model: stress-tests with galaxy clusters” supported by Next Generation EU and from the grant ASI n.\ 2024-10-HH.0 “Attività scientifiche per la missione Euclid – fase E”. MS acknowledges financial contributions from contract ASI-INAF n.2017-14-H.0, contract INAF mainstream project 1.05.01.86.10, INAF Theory Grant 2023: Gravitational lensing detection of matter distribution at galaxy cluster boundaries and beyond (1.05.23.06.17), and contract Prin-MUR 2022 supported by Next Generation EU (n.20227RNLY3, The concordance cosmological model: stress-tests with galaxy clusters). GC acknowledges the support from the Next Generation EU funds within the National Recovery and Resilience Plan (PNRR), Mission 4 - Education and Research, Component 2 - From Research to Business (M4C2), Investment Line 3.1 - Strengthening and creation of Research Infrastructures, Project IR0000012 – “CTA+ - Cherenkov Telescope Array Plus”. HH is supported by a DFG Heisenberg grant (Hi 1495/5-1), the DFG Collaborative Research Center SFB1491, an ERC Consolidator Grant (No. 770935), and the DLR project 50QE2305.
\end{acknowledgements}

\bibliography{aanda}

\appendix

\section{Halo miscentring model}\label{app:g_off}
To model the contribution due to the miscentred population of clusters, we assumed that the probability of a lens being at a projected distance $R_{\rm s}$ from the chosen centre, namely $P(R_{\text{s}})$, follows a Rayleigh distribution \citep[following e.g.][]{Johnston07,Viola15}, that is, 
\begin{equation}\label{eq:Poff}
P(R_{\text{s}}) = \frac {R_{\text{s}}} {\sigma_{\text{off}}^2} \exp \bigg[-\frac 1 2 \bigg( \frac {R_{\text{s}}} {\sigma_{\text{off}}} \bigg)^2 \bigg]\,,
\end{equation}
where $\sigma_{\rm off}$ is the standard deviation of the halo misplacement distribution on the plane of the sky, expressed in units of $h^{-1}$Mpc. The corresponding azimuthally averaged profile is given by \citep{Yang06}
\begin{equation}
\Sigma (R | R_{\text{s}}) = \frac 1 {2 \pi} \int_0^{2 \pi} \Sigma_{\text{cen}} \, \bigg( \sqrt {R^2 + R_{\text{s}}^2 - 2 R R_{\text{s}} \cos \theta} \bigg) \, {\rm d} \theta\,,
\end{equation}
where $\Sigma_{\text{cen}}$ is the centred surface mass density, while $\theta$ is the polar angle in the plane of the sky. The surface mass density distribution of a miscentred halo is expressed as
\begin{equation}\label{eq:Sigma_off}
\Sigma_{\text{off}}(R) = \int P(R_{\text{s}})\, \Sigma (R | R_{\text{s}})\, {\rm d} R_{\text{s}} \,.
\end{equation}
Analogously to Eq.\ \eqref{eq:g}, the miscentred tangential reduced shear, $g_{\rm t,\rm off}$, has the following expression:
\begin{equation}\label{eq:g_off}
g_{\rm t,\rm off}(R,M,z) = \frac{\Delta\Sigma_{\rm t,\rm off}(R,M,z)\,\langle\Sigma_{\rm crit}^{-1}(z)\rangle}{1 - \Sigma_{\rm off}(R,M,z)\,\langle\Sigma_{\rm crit}^{-1}(z)\rangle^{-1}\,\langle\Sigma_{\rm crit}^{-2}(z)\rangle}\,,
\end{equation}
where $\Sigma_{\rm off}$ is given by Eq.\ \eqref{eq:Sigma_off}, while $\Delta\Sigma_{\rm t,\rm off}$ is derived by replacing $\Sigma$ with $\Sigma_{\rm off}$ in Eq.\ \eqref{eq:DeltaSigma_cen}.

\section{Cluster--galaxy correlation model}\label{app:w_model}
The 3D halo-matter correlation function, $\xi_{\rm hm}(\vec{r})$, is defined as follows
\begin{equation}\label{eq:xi}
\xi_{\rm hm}(\vec{r}) \equiv \langle \delta_{\rm h}(\vec{x})\,\delta_{\rm m}(\vec{x}+\vec{r}) \rangle_{\vec{x}}\,,
\end{equation}
where $\vec{x}$ denotes the 3D comoving position of a halo centre, $\delta_{\rm h}$ and $\delta_{\rm m}$ are the halo and matter overdensity fields, respectively, while $\langle\dots\rangle_{\vec{x}}$ represents the average over halo positions. The matter density field is defined as
\begin{equation}\label{eq:delta_m}
\delta_{\rm m}(\vec{x}) \equiv \frac{\rho(\vec{x})-\rho_{\rm m}}{\rho_{\rm m}}\,,
\end{equation}
where $\rho$ is the local 3D matter density field and $\rho_{\rm m}$ is the mean matter density of the Universe. Because the right-hand side of Eq.~\eqref{eq:xi} involves an average over halo centres, it can be expressed in terms of the conditional matter density profile around haloes, $\rho(\vec{r}\,|\,\vec{x})$: 
\begin{equation}
\xi_{\rm hm}(\vec{r}) = \frac{\langle \rho(\vec{r}\,|\,\vec{x}) - \rho_{\rm m} \rangle_{\vec{x}}}{\rho_{\rm m}} = \frac{\langle \Delta\rho(\vec{r}\,|\,\vec{x}) \rangle_{\vec{x}}}{\rho_{\rm m}}\,.
\end{equation}
Here, $\langle\Delta\rho\rangle$ can be expressed as the \citetalias{Diemer14} profile in Eq.\ \eqref{eq:DK14}. Assuming spatial isotropy, the correlation function depends only on the separation $r=|\vec{r}|$. Furthermore, assuming that $\delta_{\rm c}=\delta_{\rm h}$, where $\delta_{\rm c}$ is the galaxy cluster density field, we can express the 3D cluster-matter correlation function as
\begin{equation}\label{eq:xi_cm}
\xi_{\rm cm}(r) = \frac{\Delta\rho(r)}{\rho_{\rm m}}\,,
\end{equation}
where we write $\langle\Delta\rho(r\,|\,x)\rangle_{x}$ as $\Delta\rho(r)$ for brevity. If galaxies are taken as biased tracers of the matter field, with galaxy overdensity $\delta_{\rm g}\equiv b_{\rm g}\delta_{\rm m}$ and galaxy bias $b_{\rm g}$, 
the corresponding cluster--galaxy correlation function follows  from Eqs.\ \eqref{eq:delta_m} -- \eqref{eq:xi_cm}:
\begin{equation}\label{eq:test0}
\xi_{\rm cg}(r) = b_{\rm g} \frac{\Delta\rho(r)}{\rho_{\rm m}}\,.
\end{equation}
Here, $b_{\rm g}$ generally depends on the scale $r$. However, as discussed below, we assumed a constant galaxy bias. The cluster--galaxy correlation function projected over the line of sight, $w_{\rm cg}$, is expressed as follows:
\begin{equation}\label{eq:test1}
w_{\rm cg}(R) = \int_0^\infty{\rm d}\chi\,\tilde{n}(\chi)\,\xi_{\rm cg}\left(\sqrt{R^2+(\chi-\chi_{\rm c})^2}\right)\,,
\end{equation}
where $R$ is the projected comoving distance from the cluster centre, $\tilde{n}(\chi)$ is the probability density function of galaxy comoving distances, and $\chi_{\rm c}$ is the comoving distance of the cluster. We remark that $\chi$ and $\chi_{\rm c}$ are comoving distances along the line of sight. Alternatively, Eq.\ \ref{eq:test1} can be integrated over the galaxy redshift, $z_{\rm g}$:
\begin{equation}\label{eq:test2}
w_{\rm cg}(R) = \int_0^\infty{\rm d}z_{\rm g}\,n(z_{\rm g})\,\xi_{\rm cg}\left(\sqrt{R^2+[\chi(z_{\rm g})-\chi_{\rm c}]^2}\right)\,,
\end{equation}
where $n(z_{\rm g})$ is the galaxy redshift probability density function. In Eq.\ \eqref{eq:test2}, the Jacobian arising from the change of variable is absorbed into the definition of $n(z_{\rm g})$, namely
\begin{equation}
n(z_{\rm g}) \equiv \tilde{n}(\chi(z_{\rm g}))\,\frac{{\rm d}\chi}{{\rm d}z_{\rm g}}\,.
\end{equation}
Substituting Eq.\ \eqref{eq:test0} into Eq.\ \eqref{eq:test2}, we obtain
\begin{equation}\label{eq:test3}
w_{\rm cg}(R) = \frac{\langle b_{\rm g}\rangle}{\rho_{\rm m}} \int_0^\infty{\rm d}z_{\rm g}\,n(z_{\rm g})\,\Delta\rho\left(\sqrt{R^2+[\chi(z_{\rm g})-\chi_{\rm c}]^2}\right)\,,
\end{equation}
where $\langle b_{\rm g}\rangle$ is the galaxy bias averaged along the line of sight. With this approximation, we assume that $b_{\rm g}$ varies slowly along the line of sight, and that $b_{\rm g}$ does not depend on the projected separation $R$. These assumptions are motivated by the limited constraining power of our data on $b_{\rm g}$.\\
\indent At this point in our analysis, we assumed that $\vec{x}$, $\vec{r}$, and $R$ are comoving distances. Conversely, in our analysis we measured and modelled $w_{\rm cg}$ as a function of the physical projected separation from cluster centres. In this case, Eq.\ \eqref{eq:test3} takes the following form:
\begin{equation}
w_{\rm cg}(R,z) = \frac{\langle b_{\rm g}(z)\rangle}{\rho_{\rm m}(z)} \int_0^\infty{\rm d}z_{\rm g}\,n(z_{\rm g})\,\Delta\rho\left(\sqrt{R^2+\left[\frac{\chi(z_{\rm g})-\chi_{\rm c}}{1+z}\right]^2},z\right)\,,
\end{equation}
where $z$ is the cluster redshift. The $(1+z)$ factor in the denominator accounts for the conversion from comoving to physical line-of-sight separation between the cluster and galaxies at the epoch of the cluster. 

\section{Robustness of the results}\label{app:robustness}
\begin{figure}[t]
\centering\includegraphics[width = \hsize-1.5cm, height = 9.cm] {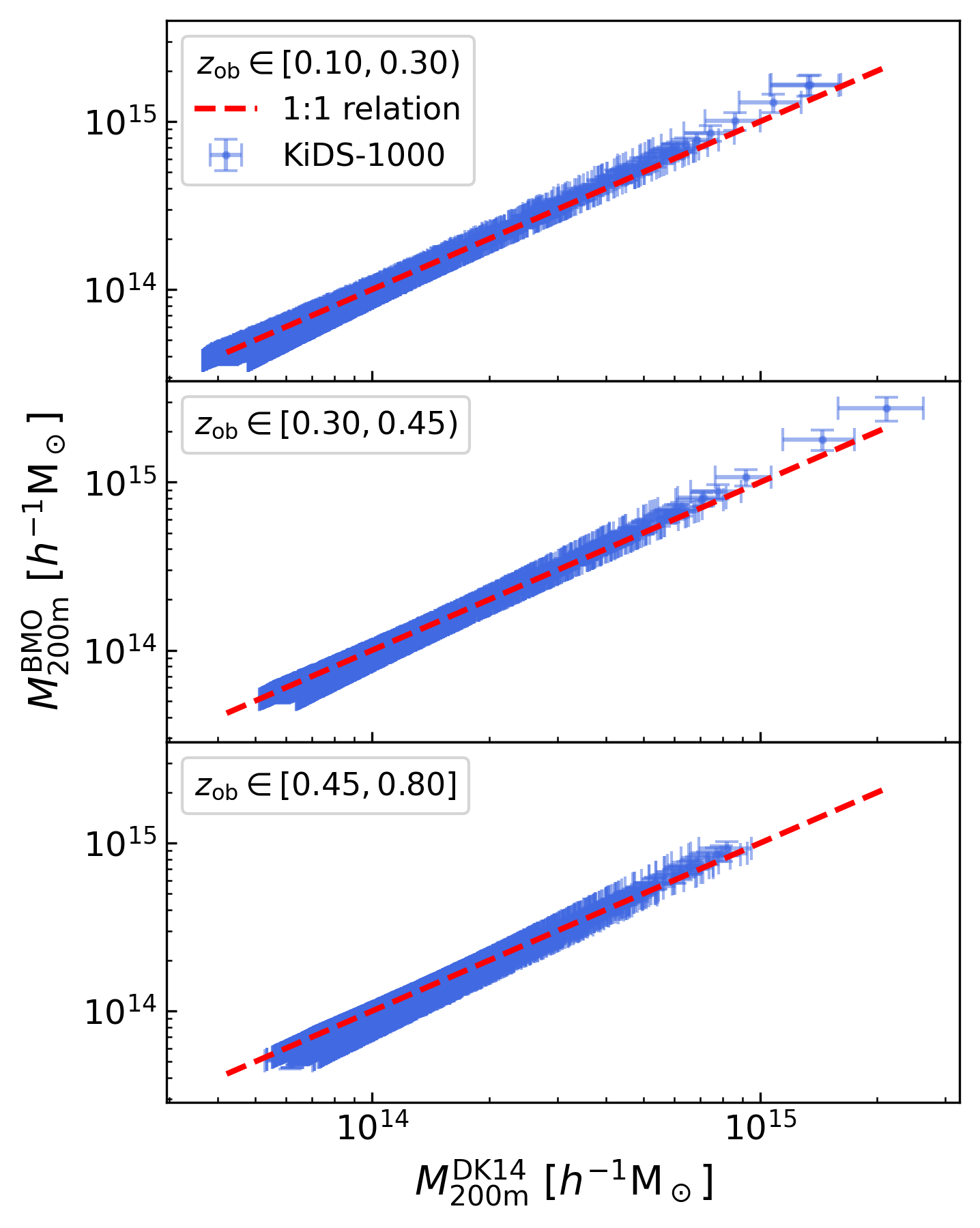}
\caption{Comparison between $M_{200}^{\rm BMO}$, derived by \citetalias{Lesci25}, and the mass estimates obtained from the $g_{\rm t}$ modelling presented in this work, $M_{200}^{\rm DK14}$, for $z_{\rm ob}\in[0.1,0.3)$ (\textit{top}), $z_{\rm ob}\in[0.3,0.45)$ (\textit{middle}), and $z_{\rm ob}\in[0.45,0.8]$ (\textit{bottom}), applying the $\lambda^*$ cuts listed in Table \ref{tab:sample} to the AMICO KiDS-1000 cluster sample. The mean and error bars are derived by marginalising the mass estimates over all the free model parameter posteriors. The dashed red lines represent the 1:1 relation.}
\label{fig:mass_comparison}
\end{figure}
\begin{figure}[t]
\centering\includegraphics[width = \hsize-1.3cm, height = 4.5cm] {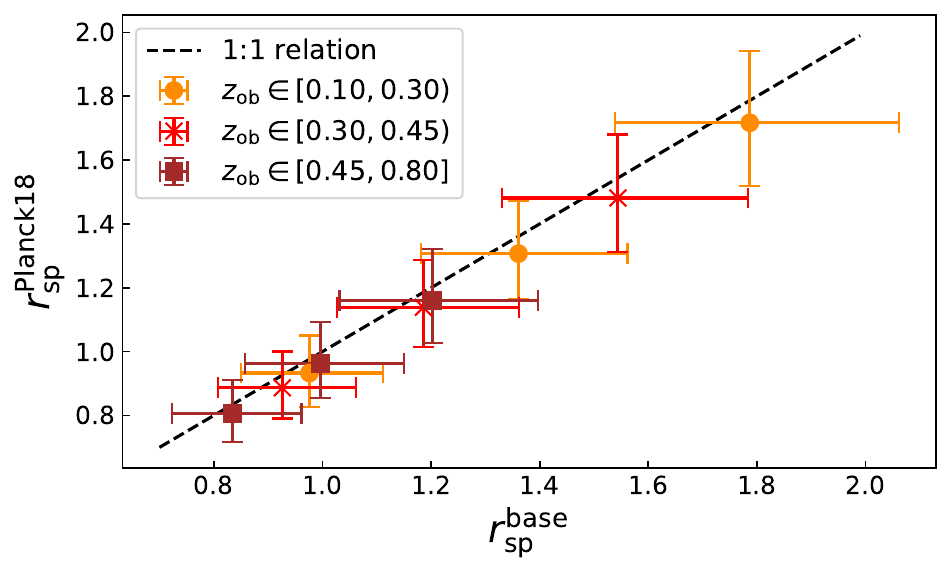}
\caption{Splashback radius derived from the $g_{\rm t}$ modelling assuming \citetalias{Planck18} cosmology, $r_{\rm sp}^{\rm Planck18}$, compared to the one derived by assuming the cosmological parameters from \citetalias{Lesci25}, $r_{\rm sp}^{\rm base}$, for $z_{\rm ob}\in[0.1,0.3)$ (orange dots), $z_{\rm ob}\in[0.3,0.45)$ (red crosses), and $z_{\rm ob}\in[0.45,0.8]$ (brown squares).}
\label{fig:rsp_planck18}
\end{figure}
\begin{figure}[t]
\centering\includegraphics[width = \hsize, height = 6.5cm] {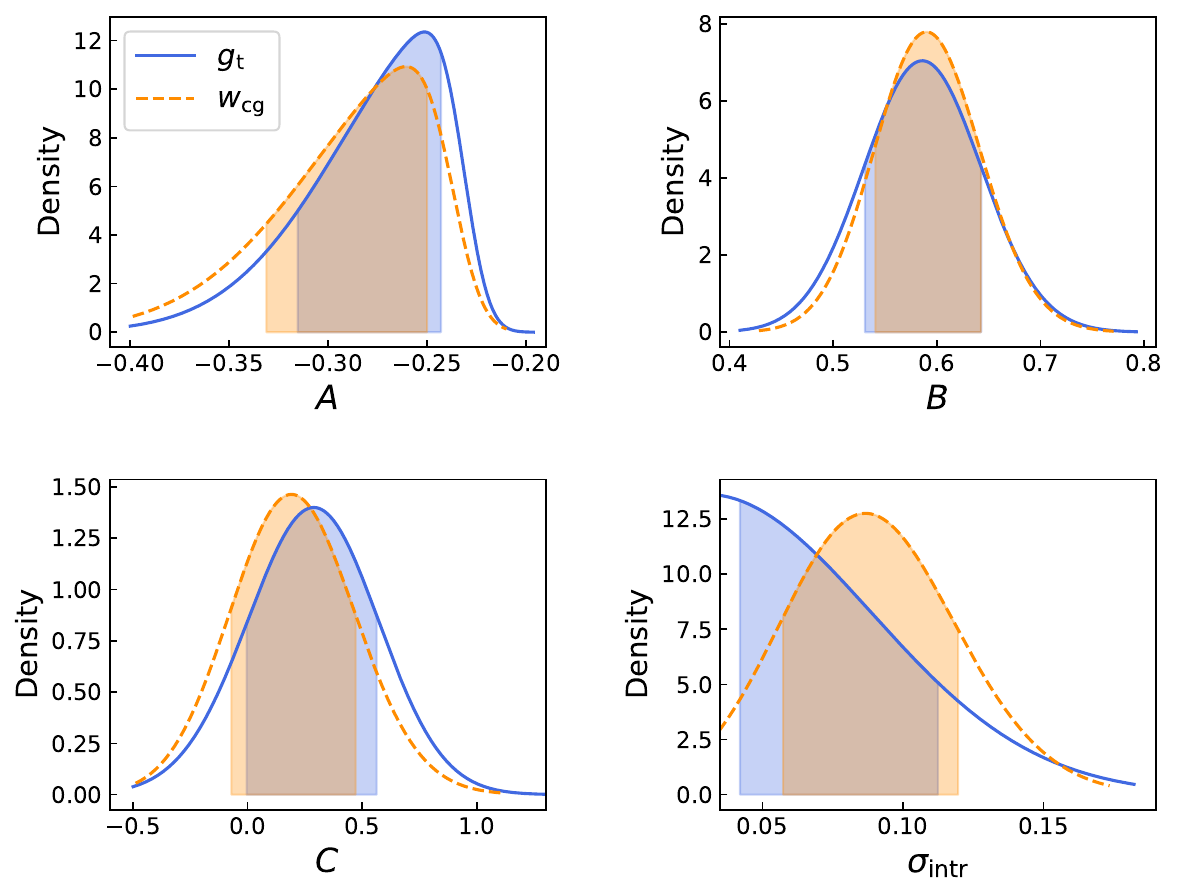}
\caption{Posterior distributions of the $\log\lambda^*-\log M_{200\rm m}$ relation parameters (Eq.\ \ref{eq:scalingrelation_PDF}) from the modelling of $g_{\rm t}$ (solid blue lines) and $w_{\rm cg}$ (dashed orange lines). Shaded areas represent 68\% confidence regions.}
\label{fig:scaling}
\end{figure}
\begin{figure}[t]
\centering\includegraphics[width = \hsize-1.2cm, height = 4.5cm] {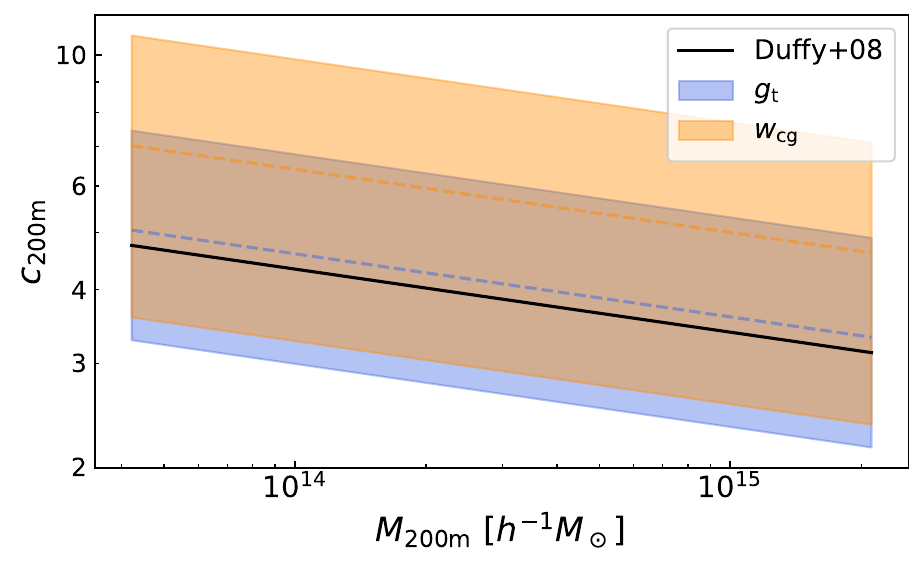}
\caption{$\log c_{200\rm m}-\log M_{200\rm }$ relation constrained by $g_{\rm t}$ (blue band) and $w_{\rm cg}$ (orange band) measurements. The width of the bands represents the 68\% confidence of the models, while the dashed lines mark median values. The solid black line shows the model from \citet{Duffy08}. The cluster redshift assumed in this figure is $z=0.45$.}
\label{fig:concentration}
\end{figure}
\begin{figure*}[t]
\centering\includegraphics[width = \hsize-2.7cm, height = 14.5cm] {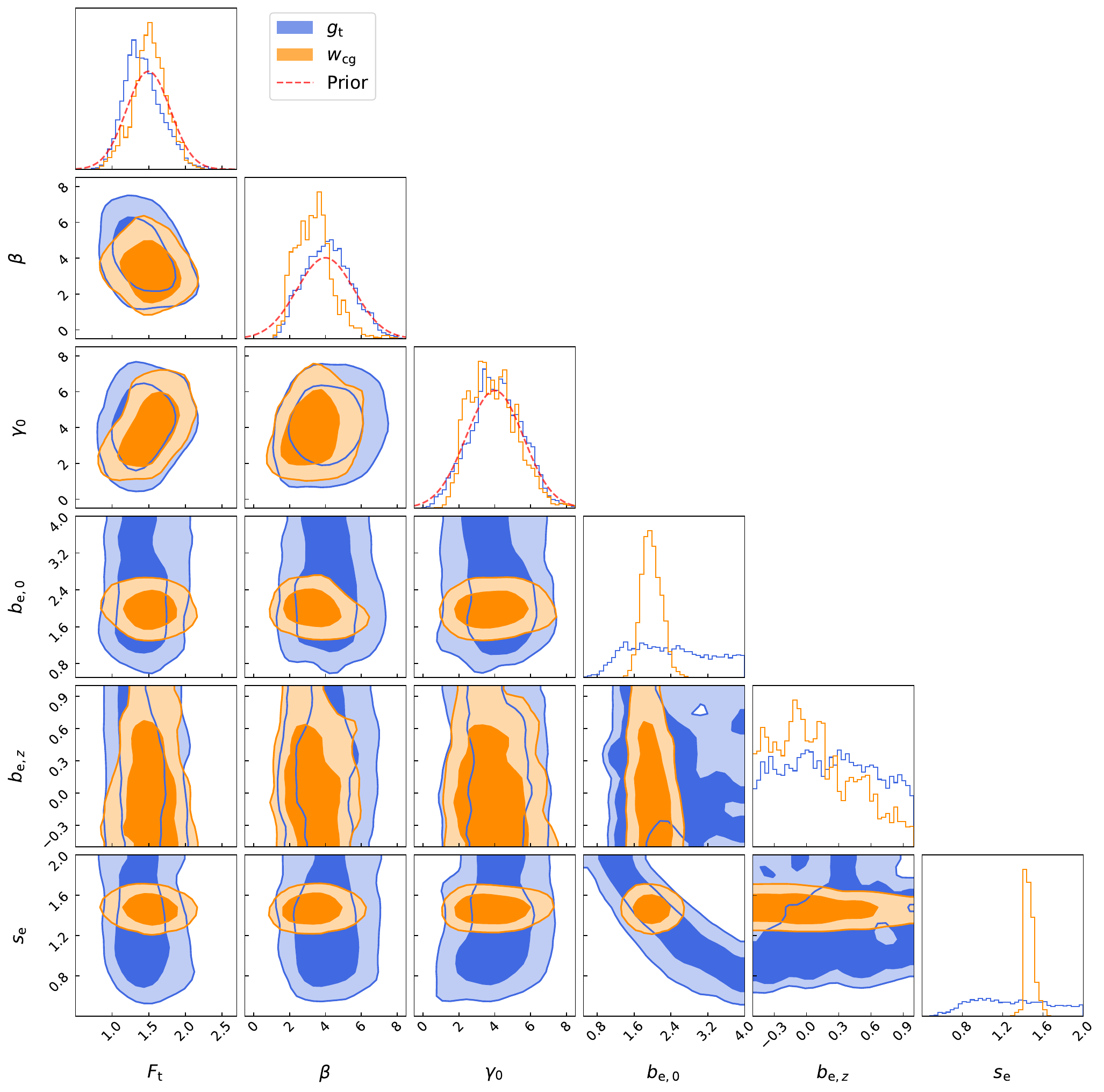}
\caption{Constraints on the parameters of the transition factor and infalling term of the \citetalias{Diemer14} profile. The 68\% and 84\% confidence levels of the 2D posteriors from $g_{\rm t}$ (blue) and $w_{\rm cg}$ (orange) modellings are shown. The Gaussian priors on $F_{\rm t}$, $\beta$, and $\gamma_0$ are represented by the dashed red lines.}
\label{fig:DK14_parameters}
\end{figure*}
In this work parameters that cannot be fully constrained by our data are fixed using simulations and previous observational results, as described in Sect.\ \ref{sec:modelling:likelihood}. Moreover, our sample selections are driven exclusively by the quality of the fit. For instance, we excluded clusters at $z>0.8$ since their inclusion would otherwise reduce the fit accuracy. In this section we assess the impact of these assumptions on our final results.\\
\indent Several tests were already performed in \citetalias{Lesci25}, who carried out the weak-lensing mass calibration of the cluster sample analysed in this paper. They showed that weak-lensing measurements pass null tests, further demonstrating that the final mass constraints are robust against the choice of priors on the halo miscentring parameters. The same analysis also confirmed that our data lack the precision to detect any potential mass dependence of $\sigma_{\rm intr}$. In addition, \citetalias{Lesci25} demonstrated that fixing the mass slope of the $\log c_{200\rm m}-\log M_{200\rm m}$ relation (Eq.\ \ref{eq:cM_rel}), denoted as $c_M$ and discussed in Sect.\ \ref{sec:modelling:likelihood}, does not introduce any biases in the mass estimates. The same holds for the intrinsic scatter in the $\log c_{200\rm m}-\log M_{200\rm m}$ relation, which amounts to about 35\% \citep[see e.g.][]{Duffy08,Bhattacharya13,Diemer15}. Given our dataset, this scatter is negligible compared to the statistical uncertainty on the amplitude of the relation, $\log c_0$.\\
\indent Figure \ref{fig:mass_comparison} displays the comparison between the individual cluster mass estimates derived from the $g_{\rm t}$ modelling presented in this work (based on Eq. 54 in \citetalias{Lesci25}) and those from \citetalias{Lesci25}. Despite \citetalias{Lesci25} adopting the truncated NFW profile by \citet*[][BMO]{BMO} rather than the \citetalias{Diemer14} profile in Eq. \eqref{eq:DK14}, and modelling the $g_{\rm t}$ profiles only up to $3.5$ $h^{-1}$Mpc, we find excellent agreement. \\
\indent As discussed in Sect.\ \ref{sec:results}, in our analysis we assumed the median cosmological parameter values constrained by \citetalias{Lesci25}. To test the impact of this choice, we fixed the cosmological parameters to the median values from \citetalias{Planck18}. The resulting splashback radius estimates obtained from the $g_{\rm t}$ modelling are systematically lower but consistent within $0.5\sigma$ (Fig.\ \ref{fig:rsp_planck18}). These lower $r_{\rm sp}$ estimates are expected, as the $\Omega_{\rm m}$ value from \citetalias{Planck18}, $\Omega_{\rm m}=0.31$, is larger than the one assumed in our baseline analysis, analysis, namely $\Omega_{\rm m}=0.22$. Indeed, larger $\Omega_{\rm m}$ implies faster accretion rates and smaller $r_{\rm sp}$ \citep{Diemer17,Haggar24,Mpetha24}. We verified that the same level of agreement holds for $\mathcal{R}_{\rm sp}$. While this test was performed only for $g_{\rm t}$ modelling, we expect similar results for $w_{\rm cg}$.

\section{Constraints on the DK14 parameters}\label{app:nuisance}
In this work we first estimated the parameters of the $\log\lambda^*-\log M_{200\rm m}$ relation (Eq.\ \ref{eq:scalingrelation_PDF}) by modelling $g_{\rm t}$, subsequently using the resulting posteriors as priors for the $w_{\rm cg}$ modelling. Figure \ref{fig:scaling} shows that the two probes yield similar constraints on these parameters. The most significant differences appear in the posteriors for the amplitude, $A$, and the intrinsic scatter, $\sigma_{\rm intr}$. For $w_{\rm cg}$, the $A$ posterior is shifted to lower values and is more heavily tailed, while $\sigma_{\rm intr}$ peaks at a larger value compared to the constraint from $g_{\rm t}$. The $\sigma_{\rm intr}$ result can be explained by our assumption of $M_{200\rm m} \equiv M_{g_{\rm t}} \equiv M_{w_{\rm cg}}$, where $M_{g_{\rm t}}$ and $M_{w_{\rm cg}}$ are the weak-lensing and cluster--galaxy correlation function masses, respectively. In fact, these are scattered proxies of the true $M_{200\rm m}$ \citep[see e.g.][]{Sereno15_I}, with a larger scatter expected for $M_{w_{\rm cg}}$ due to the large variance in cluster member galaxy distributions. The differences in $A$ and $\sigma_{\rm intr}$ explain the larger statistical uncertainties on $M_{w_{\rm cg}}$ compared to $M_{g_{\rm t}}$ (Fig.\ \ref{fig:precision}). \\
\indent Figure \ref{fig:concentration} shows that our constraints on the $\log c_{200\rm m}-\log M_{200\rm m}$ relation agree within 1$\sigma$ with the model by \citet{Duffy08}. The $r_{\rm sp}$ results presented in Sect.\ \ref{sec:results} suggest that dynamical friction on cluster member galaxies imprints a splashback feature in $w_{\rm cg}$ at smaller scales than in weak-lensing observations. Consequently, haloes traced by cluster galaxies should appear more concentrated. However, the large uncertainty on $c_{200\rm m}$ derived from $w_{\rm cg}$ (Fig.\ \ref{fig:concentration}) precludes any stronger conclusion on this point.\\
\indent Figure \ref{fig:DK14_parameters} shows that the constraints on the parameters of the \citetalias{Diemer14} transition factor (Eq.\ \ref{eq:DK14:transition}), namely $F_{\rm t}$, $\beta$, and $\gamma_0$, align with the adopted Gaussian priors. The largest difference between prior and posterior occurs for $\beta$ in the $w_{\rm cg}$ modelling, where the posterior median shifts from the prior by $0.5\sigma$ (see also Table \ref{tab:priors_and_posteriors}). Furthermore, in this work we adopted uninformative priors on the infalling profile parameters in Eq.\ \eqref{eq:DK14:infall}, namely $b_{\rm e}$ and $s_{\rm e}$, allowing $b_{\rm e}$ to evolve with redshift (Eq.\ \ref{eq:b_e}). As shown in Fig.\ \ref{fig:DK14_parameters} and reported in Table \ref{tab:priors_and_posteriors}, $g_{\rm t}$ measurements do not constrain these parameters, whereas $w_{\rm cg}$ provides strong constraints with a precision of 11\% on $b_{{\rm e},0}$ and of 4\% on $s_{\rm e}$. Our resulting values, $b_{{\rm e},0}=1.97^{+0.23}_{-0.19}$ and $s_{\rm e}=1.461^{+0.052}_{-0.046}$, also agree within 1$\sigma$ with the theoretical predictions from \citetalias{Diemer14}. Neither probe constrains the redshift evolution of $b_{\rm e}$.  

\end{document}